\documentstyle[preprint,aps,tighten,eqsecnum]{revtex}

\begin{document}


\def \ns{\enspace}
\def \ts{\thinspace}
\def \nts{\negthinspace}
\def \beq{\begin{equation}}
\def \eeq{\end{equation}}
\def \beqa{\begin{eqnarray}}
\def \eeqa{\end{eqnarray}}
\def \eps{\epsilon}
\def \EPS{\varepsilon}
\def \trace{{\rm Tr}}
\def \half{\hbox{$1\over2$}}
\def \zero{{\bf 0}}
\def \zeroIII{\vec{0}}
 

\def \Ait#1#2{{A}^{#1}_{(#2)}}
\def \xf{x_{{}_F}}
\def \mn{m}
\def \Isng{{\cal I}_{sng}}
\def \Ismth{{\cal I}_{smth}}
\def \Kcube{K^3}
\def \aw{a_\omega}
\def \GIII{G}
\def \Lbrk{\Bl\nts\Bl}
\def \Rbrk{\Br\nts\Br}
\def \zed{{\cal Z}}
\def \LQCD{\Lambda_{\rm QCD}}
\def \Jt{\bbox{J}}
\def \kt{\bbox{k}}
\def \pt{\bbox{p}}
\def \pz{p_{\zeta}}
\def \pIII{\vec{p}}
\def \qt{\bbox{q}}
\def \qz{q_{\scriptscriptstyle\parallel}}
\def \qIII{\vec{q}}
\def \rIII{\vec{r}}
\def \dz{\partial_{\scriptscriptstyle\parallel}}
\def \xz{x_{\scriptscriptstyle\parallel}}
\def \yz{y_{\scriptscriptstyle\parallel}}
\def \uz{\Upsilon_{\scriptscriptstyle\parallel}}
\def \xiz{\xi_{\scriptscriptstyle\parallel}}
\def \xit{\bbox\xi}
\def \xiIII{\vec{\xi}}
\def \xiprimeIII{\xiIII\ts^\prime}
\def \deltat{\bbox\Delta}
\def \deltaIII{\vec\Delta}
\def \deltaz{\Delta_{\scriptscriptstyle\parallel}}
\def \sigmat{\bbox\Sigma}
\def \sigmaIII{\vec\Sigma}
\def \sigmaz{\Sigma_{\scriptscriptstyle\parallel}}
\def \delt{\bbox\nabla}
\def \xt{\bbox{x}}
\def \xIII{\vec{x}}
\def \xprimet{\xt^\prime}
\def \xprimeIII{\xIII\ts^\prime}
\def \Bl{\bbox{\Bigl[}}
\def \Br{\bbox{\Bigr]}}
\def \Qprobe{Q}
\def \curlyL{{\cal L}}
\def \curlyE{{\cal E}}
\def \vsqr{v^2}
\def \DIII{{\cal D}}
\def \SIII{{\cal S}}
\def \U{{\rm U}}
\def \Uinv{\U^{-1}}
\def \pexp{{\cal P}\exp}
\def \gnum{ {{dN}\over{dq^{+} d^2\qt}} }
\def \gnumX{ {{dN}\over{d\xf d^2\qt}} }
\def \gnumIII{ {{dN}\over{d\qz d^2\qt}} }
\def \Ihat{C}
\def \sgn{\mathop{\rm sgn}\nolimits}


\draft
\preprint{
  \parbox{2in}{McGill/00--17 \\
  hep-ph/0007133
}  }

\title{Longitudinal Resolution in a Large Relativistic Nucleus:\\
Adding a Dimension to the McLerran-Venugopalan Model}
\author{C.S. Lam~\cite{LAMemail} and Gregory Mahlon~\cite{GDMemail}}
\address{Department of Physics, McGill University, \\
3600 University Street, Montr\'eal, Qu\'ebec  H3A 2T8 \\
Canada }
\date{July 13, 2000}
\maketitle
\begin{abstract}
We extend the McLerran-Venugopalan model for the gluon distribution
functions of very large nuclei to larger values of the longitudinal 
momentum fraction $\xf$.  Because gluons with larger values of $\xf$ 
begin to resolve the longitudinal structure of the nucleus, we
find that it is necessary to set up a fully three-dimensional
formalism for performing the calculation.  We obtain a relatively 
compact expression for the gluon number density provided that the 
nucleus is sufficiently large and consists of color-neutral nucleons.
Our expressions for the gluon number density saturate at small 
transverse momenta.  The nuclear dependence we obtain is such that the
number of gluons increases more slowly than the number of nucleons
is increased.
\end{abstract}
\pacs{24.85.+p, 12.38.Bx}


\section{Introduction}

The recent construction and commissioning of the Brookhaven
Relativistic Heavy Ion Collider has led to a renewed interest
in the properties of heavy nuclei in recent years. 
A considerable amount of fruitful work has 
been done on classical and semiclassical
descriptions of the physics 
involved\cite{paper1,paper2,paper3,paper4,paper9,PAPER14,%
paper18,paper21,paper19,paper58,paper12,paper24,paper10,%
paper27,paper15,paper41,paper37,paper59,paper48}.
In particular,
the McLerran-Venugopalan (MV) 
model\cite{paper1,paper2,paper3,paper4,paper9}
provides a framework for calculating the gluon distribution
functions for very large nuclei at very small values of the
longitudinal momentum fraction $\xf$.
What McLerran and Venugopalan realized is that at sufficiently
small $\xf$, the gluons are unable to resolve the longitudinal
structure of the nucleus, meaning that many quarks contribute
to the color field at each value of the (transverse) position $\xt$.
This large charge per unit area $\kappa^2$ provides the
scale at which the strong coupling is evaluated\cite{paper1}.
Thus, if $\kappa^2 \gg \LQCD^2$, a classical treatment ought
to provide a reasonable description. 

Recently, we pointed out that the infrared divergences which
appear in the MV model may be cured by incorporating the
effects of confinement\cite{PAPER14}.  That is, we observe
that nucleons display no net color charge:  individual
quarks are confined inside the nucleons, whose radius
is $a \sim \LQCD^{-1}$.  
As a consequence, we
expect that there should not be long range ($\gg a$)
correlations between quarks.  Strong correlations between quarks
occur only when we probe at short distance scales.
These considerations may be phrased as a mathematical
constraint on the form of the two-point charge density correlation
function.  

The calculations presented in 
Refs.~\cite{paper1,paper2,paper3,paper4,paper9,PAPER14} all assume
$\xf$ to be small enough so that the gluons do not probe the
longitudinal structure of the Lorentz-contracted nucleus which
they see.  Effectively, then, the relevant geometry is two-dimensional,
with the source exactly on the light cone.
In this work we extend the MV model to larger values of the
longitudinal momentum fraction $\xf$.  In this regime, the
gluons begin to resolve the longitudinal structure of the nucleus:
therefore, we develop a fully three-dimensional framework using a
source that is not quite aligned with the light cone.
In order to deal with the complications which arise as a result,
we must rely heavily on the fact that we consider a large nucleus
of radius $R \gg a$ which consists of color-neutral nucleons.  

The remainder of this paper is organized as follows.  In 
Sec.~\ref{SOURCE} we present our conventions for
writing down the classical Yang-Mills equations
and describe the (slightly) off-light-cone source which will
be the foundation of our calculation.  We show that the natural
(order unity) variables to describe the nucleus are essentially
those in the nuclear rest frame, even in the limit $\beta\rightarrow 1$.
In Sec.~\ref{COUNTING} we set up the framework for determining
the gluon number density in the Weizs\"acker-Williams approximation.
In this section we introduce the two-point charge density correlation
function, and review the color-neutrality condition\cite{PAPER14} 
which it must satisfy.  Sec.~\ref{VALIDITY}
contains a discussion of the requirements which
must be satisfied in order for our approximations to be valid.
The meat of our calculation is contained in Sec.~\ref{CALCULATION},
where we begin with the solution for the vector potential
in the covariant gauge, perform the transformation to light-cone
gauge, and determine the gluon number density.  Additional
details of this calculation are found in Appendix~\ref{CONTRACTIONS}.
We illustrate our results for the gluon number density with the
help of a power-law model for the correlation function in 
Sec.~\ref{EXAMPLE}.  The integrals which arise in connection with
this model are presented in Appendix~\ref{PowerDetails}.
Finally, Sec.~\ref{CONCLUSIONS} contains our conclusions.


\section{The Classical Yang-Mills Equations and Source}\label{SOURCE}

In this section we 
present the conventions which we use in writing down the
Yang-Mills equations and source for a nucleus which moves down
the $z$ axis with a speed $\beta<1$.
The motivation behind the choices we have made is to ensure that all 
``unknown'' quantities are of order unity, with all powers of the
small and large parameters explicitly written out.  
This will make the approximations which we will have to make
later on more transparent.  It will also make the $\beta\rightarrow 1$
limit which we take at the end obvious.

We begin with the classical Yang-Mills equations, which we write
as
\beq
D_\mu F^{\mu\nu} = gJ^{\nu},
\label{YangMills}
\eeq
where we have employed matrix form, {\it i.e.}
$ J^{\nu} \equiv T^a J^{a\nu}$, etc.
The $T^a$ are the normalized Hermitian generators of SU($N_c$)
in the fundamental representation, satisfying 
$2 \ts\trace(T^a T^b) = \delta^{ab}$.
The covariant derivative is
\beq
D_\mu F^{\mu\nu} \equiv
\partial_\mu F^{\mu\nu} - ig \Bl A_\mu,F^{\mu\nu} \Br
\eeq
and the field strength reads
\beq
F^{\mu\nu} \equiv 
\partial^{\mu} A^{\nu} 
-\partial^{\nu} A^{\mu} 
-ig \Bl A^{\mu},A^{\nu} \Br.
\label{Fmunu}
\eeq
The conventions contained in Eqs.~(\ref{YangMills})--(\ref{Fmunu})
ensure that all powers of the strong coupling constant $g$ are
explicit, with no hidden $g$-dependence.

We now turn to the source appearing in~(\ref{YangMills}),
starting in the rest frame of the nucleus.
Using the subscript ``$r$'' to denote rest frame quantities,
the current takes on the simple form
\beq
J_r^0 = \rho(-z_r,\xt_r); \qquad
J_r^1 = J_r^2 = J_r^3 = 0.
\label{Jrest}
\eeq
The color charge density $\rho\equiv T^a\rho^a$ is a spherically 
symmetric function which is non-zero over a region of size $R$,
the radius of the nucleus.  
Since in the lab frame we want the nucleus to be moving along the
$+z$ axis, it is convenient to use $-z_r$ for the longitudinal
coordinate in Eq.~(\ref{Jrest}).  The transverse coordinates
$x_r$ and $y_r$ form a two-vector which we write in bold-face: $\xt_r$.
In terms of the light-cone coordinates\footnote{
Our metric has the signature $({-},{+},{+},{+})$.  Thus,
the scalar product in light-cone coordinates reads
$ q^{\mu} x_{\mu} = -q^{+}x^{-} - q^{-}x^{+} + \qt\cdot\xt. $
We will think of $x^{+}$ as the time,
and $x^{-}$ as the longitudinal distance.
}
$x^{\pm} = -x_{\mp} = (x^0 \pm x^3)/\sqrt{2},$
Eq.~(\ref{Jrest}) may be written as
\beq
J^{+}_r = J^{-}_r = \hbox{${1}\over{\sqrt{2}}$} 
\rho\Bigl( \hbox{${1}\over{\sqrt{2}}$}(x^{-}_r{-}x^{+}_r),\xt_r\Bigr); 
\qquad \Jt_r = \zero.
\label{JrestLC}
\eeq
The net color charge of the nucleus is zero:
\beq
\int dz_r \ts d^2\xt_r \ts\ts \rho(-z_r,\xt_r) = 0.
\eeq
The nucleus is not a homogeneous sphere of color charge: it
has substructure.
Because of confinement, there are smaller regions of size 
$a\sim\LQCD^{-1}$
within the volume occupied by the nucleus for which the total color
charge also vanishes.  These regions correspond to the nucleons.
For a large nucleus, $a/R \approx A^{-1/3} \ll 1$.

Boosting to the lab frame, where the nucleus moves along the $+z$
axis with a speed $\beta$, Eq.~(\ref{JrestLC}) becomes
\beq
J^{+} = {{1}\over{\EPS}} \ts 
\rho\Bigl( \hbox{${{1}\over{\EPS}}$}x^{-}  {-}
           \hbox{${{\EPS}\over{2}}$}x^{+}  ,\xt\Bigr); 
\qquad J^{-} = {{\EPS}\over{2}} J^{+};
\qquad \Jt = \zero,
\label{Jlab}
\eeq
where we have defined
\beq
\EPS \equiv \sqrt{ { {2(1-\beta)} \over {1+\beta} }}.
\label{epsdef}
\eeq
Viewed in the lab frame, the nucleus is Lorentz-contracted
to a thickness of order $R/\gamma$, where, as usual,
$\gamma \equiv (1{-}\beta^2)^{-1/2}$.
For large boosts
the charge density function is non-zero only when $x^{-} \sim \EPS R$.
Hence, the longitudinal argument in Eq.~(\ref{Jlab}) is really 
of order $R$, leading us to define the new longitudinal
variable
\beq
\xz \equiv 
{{1}\over{\EPS}}\ts x^{-} 
- {{\EPS}\over{2}}\ts  x^{+}.
\label{xl-def}
\eeq
The current appearing in Eq.~(\ref{Jlab}) is a function of
$(\xz,\xt) \equiv \xIII$.  The use of the notation ``$\xIII$\ts'' 
is suggestive of the fact that, in terms of the natural (order unity)
variables, the functions describing the nucleus are 
{\it still spherical!}   In fact, $\xz$ is just the (unboosted)
longitudinal coordinate from the rest frame.  The advantage of
using $\xz$ instead of $x^{-}$ should be obvious:  to take
the $\beta\rightarrow 1$ ($\EPS\rightarrow 0$) limit for quantities
written in terms of $\xz$ is trivial, whereas if the same quantities
were written in terms of $x^{-}$ instead, we would have to be careful
to hold $x^{-}/\EPS$ fixed.

Because the choice made in~(\ref{Jrest}),
we define
\beq
\qIII \cdot \xIII \equiv
-\qz\xz + \qt\cdot\xt.
\label{dot3}
\eeq
We will also use the notation $d^3\xIII \equiv d\xz\ts d^2\xt$.
All of the functions we will be dealing with will depend only
on the particular combination of $x^{+}$ and $x^{-}$ appearing
in Eq.~(\ref{xl-def}).  Thus, we have the replacements
\beq
\partial_{+} \rightarrow -{{\EPS}\over{2}} \ts {{\partial}\over{\partial\xz}}
\equiv -{{\EPS}\over{2}} \ts \dz ; \qquad
\partial_{-} \rightarrow {{1}\over{\EPS}} \ts \dz .
\label{x1594-x}
\eeq

The divergence of the current in Eq.~(\ref{Jlab}) vanishes:
$\partial_{\mu} J^{\mu} = 0$.
In QCD, however, we require that the current be covariantly
conserved: 
$D_{\mu} J^{\mu} = 0$.
Since we are at weak coupling, we may work iteratively.
That is, we first solve the Yang-Mills equations using~(\ref{Jlab})
for the source.  The resulting solution will violate the
covariant conservation condition by an amount of order $g^2$.
The value of $D_{\mu}J^{\mu}$ which is obtained could, in
principle, be used to correct the current to this order
given the non-Abelian equivalent of the Lorentz force equation
to provide information on how the color of the quarks making up
the source changes upon emission of a gluon.
Since we are assuming that $g^2 \ll 1$, we will simply drop
these contributions.

As suggested by Eq.~(\ref{dot3}), we will use the notation $\qz$
for the component of momentum conjugate to $\xz$.
To make the connection with the longitudinal momentum fraction $\xf$,
let the gluon carry a momentum $q^{+}$ and the nucleon a momentum
$Q^{+}$.  Then,
\beq
\xf \equiv  
{ { q^{+} } \over { Q^{+} } }
= { {\EPS q^{+}}\over{\mn} },
\label{x1596a}
\eeq
where $\mn$ is the nucleon mass.  
However, 
\beq
q^{+} \leftrightarrow i\partial_{-} 
      \leftrightarrow {{i}\over{\EPS}}\ts \dz
      \leftrightarrow {{\qz}\over{\EPS}}.
\eeq
Thus, we conclude that
\beq
\xf \equiv {{\qz}\over{\mn}}.
\label{XFdef}
\eeq
Once more note the advantage of the rest-frame variables over
the light-cone variables:  in terms of $q^{+}$, we would have
to take the $\EPS\rightarrow 0$ limit with the caveat that the
combination
$\EPS q^{+}$ is held fixed.  No such complication arises when
we use $\qz$ instead.


\section{Counting Gluons}\label{COUNTING}

Next, we turn to the formula for the gluon number density.
Recall that the standard expression
reads\cite{CollinsSoper}
\beq
\gnum = 
{ {q^{+}}\over{4\pi^3} } \nts
\int_{-\infty}^{\infty} \nts dx^{-}
\int_{-\infty}^{\infty} \nts dx^{\prime -}
\int d^2\xt
\int d^2\xprimet \ts
e^{-iq^{+}(x^{-}-x^{\prime -})}
e^{i\qt\cdot(\xt-\xprimet)}
\langle A_i^a(x^{-},\xt) A_i^a(x^{\prime -},\xprimet) \rangle.
\label{gnum-standard}
\eeq
Eq.~(\ref{gnum-standard}) is written in terms of the light-cone
variables $x^{-}$ and $q^{+}$, and the light-cone gauge vector
potential $A_i$.  
The light-cone gauge has favored status with respect to the
intuitive picture of the parton 
model\cite{CollinsSoper,Curci,LC1,paper31}: thus we
continue to use the light-cone gauge even for a source which
moves at less than the speed of light.  
Based on the discussion of Sec.~\ref{SOURCE}, however,
we wish to employ
the ``new'' longitudinal variables $\xz$ and $\qz$; therefore,
we write
\beqa
\gnumIII &\equiv &
{ {\qz}\over{4\pi^3} } \ts
\int d^3\xIII
\int d^3\xIII\ts' \ts
e^{i\qIII\cdot(\xIII-\xprimeIII)}
\langle A_i^a(\xIII\ts) A_i^a(\xIII\ts') \rangle
\nts\nts\nts\nts
\cr &=&
{ {\qz}\over{4\pi^3} } \ts
\int d^2\xt
\int d^2\xprimet \ts
e^{i\qt\cdot(\xt-\xprimet)}
\langle A_i^a(\qz;\xt) A_i^a(-\qz;\xprimet) \rangle,
\label{gnum3d}
\eeqa
where
\beq
A(\qz;\xt) \equiv
\int_{-\infty}^{\infty} d\xz\ts e^{-i\qz\xz} \ts A(\xIII\ts).
\eeq
In the limit $\EPS\rightarrow 0$, Eq.~(\ref{gnum3d}) reduces
to the previous result, Eq.~(\ref{gnum-standard}).

Eq.~(\ref{gnum3d}) produces a gluon number density which
is differential not only in $\qz$, but in the transverse
momentum as well.
To obtain the usual gluon structure function resolved at the scale
$\Qprobe^2$, we simply supply the trivial factor of $\mn$ required
to convert $\qz$ into $\xf$ and integrate~(\ref{gnum3d})
over all transverse momenta less than or equal to $\Qprobe$:
\beq
g_A(\xf,\Qprobe^2)  \equiv 
\int_{\vert\qt\vert\le\Qprobe}\nts\nts\nts  d^2\qt \ts
{{dN}\over{d\xf d^2\qt}}.
\label{StrFnDef}
\eeq

Our classical approximation to the quantum average represented
by the angled brackets appearing on the right-hand-side of 
Eq.~(\ref{gnum3d}) consists of performing an
ensemble average with a Gaussian weight.
We parameterize the two-point charge density correlation function 
by
\beq
\langle \rho^a(\xIII\ts) \rho^b(\xIII\ts') \rangle \equiv
\delta^{ab} \ts \Kcube
{\SIII}\biggl( {{\xIII+\xIII\ts'}\over{2}} \biggr)
{\DIII}(\xIII-\xIII\ts').
\label{rhorho3}
\eeq
The functions $\SIII$ and $\DIII$ appearing in this definition 
encode two different aspects of the physics of the nucleus.
The color-neutrality condition developed in Ref.~\cite{PAPER14}
imposes the following constraint on~$\DIII$:
\beq
\int d^3\xIII \ts\ts\DIII(\xIII\ts) = 0.
\label{CNpos}
\eeq
In terms of the Fourier-transformed function $\widetilde\DIII$,
this constraint reads
\beq
\widetilde\DIII (0,\zero) = 0.
\label{CNmom}
\eeq
When the color-neutrality condition is satisfied,
the function $\DIII$ contains an intrinsic scale, reflecting the
minimum size of the region for which Eq.~(\ref{CNpos}) is
approximately true.  Because of confinement, we expect this scale
to be roughly the nucleon radius $a \sim \LQCD^{-1}$,
reflecting the fact that points within different nucleons ought
to be (largely) uncorrelated.
On the other hand, the function 
$\SIII\Bigl(\half(\xIII{+}\xIII\ts')\Bigr)$, 
which depends on the center-of-mass coordinate, 
should be non-negligible over
a region of size $R \sim A^{1/3} a$, the radius of the entire nucleus.
We choose to normalize $\SIII$ so that its total integral
simply gives the volume of the nucleus:
\beq
\int d^3\sigmaIII \ts\ts \SIII(\sigmaIII\ts) \equiv V.
\label{Snorm}
\eeq
We have written $V$ in Eq.~(\ref{Snorm})
rather than the ${4\over3} \pi R^3$
pertaining to a spherical nucleus to maintain generality
and to aid in making connection to Refs.~\cite{paper9,PAPER14}.
The detailed forms of the functions $\SIII$ and $\DIII$ depend upon 
aspects of non-perturbative QCD which are
poorly understood.  However, the requirements specified above
account for the relevant physics:
we have a correlation function which takes on non-trivial values
only for points which are close enough together to lie
inside a single nucleon ($\vert\xIII-\xIII\ts'\vert \alt a$)
and which are centered anywhere inside the nucleus
($\vert\half(\xIII+\xIII\ts')\vert \alt R$).
The interplay
between $\SIII$ and $\DIII$ will be crucial in helping us to
organize our calculation in powers of $a/R$.

In addition to satisfying the color neutrality condition,
$\DIII(\xIII-\xprimeIII)$ should contain a term which goes like 
$\delta^3(\xIII-\xprimeIII)$\cite{paper1,paper2,paper3,paper4,paper9,paper12}.
The presence of such
a contribution leads to a point-like $1/\qt^2$ behaviour in
the gluon number density at large $\qt^2$\cite{PAPER14}, consistent
with the physics of asymptotic freedom.  
This term arises from the self-correlation
of the quarks.
Quite generally, then, we expect the form of
$\DIII(\xIII-\xprimeIII)$ to be
\beq
\DIII(\xIII-\xprimeIII) = 
\delta^3(\xIII-\xprimeIII) - \Ihat(\xIII-\xprimeIII),
\label{TwoTerms}
\eeq
where $\Ihat(\xIII-\xprimeIII)$ is a
reasonably smooth function parameterizing the
mutual correlations between pairs of quarks.    
Because $\Ihat(\xIII-\xprimeIII)$ describes the structure of a 
color-neutral nucleon, it must have unit integral and
possess non-trivial values only when
$\vert\xIII-\xprimeIII\vert \alt a \sim \LQCD^{-1}$.
In momentum space, Eq.~(\ref{TwoTerms}) reads
\beq
\widetilde\DIII(\qIII\ts) = 1 - \widetilde\Ihat(\qIII\ts).
\label{TwoTermsII}
\eeq
The color neutrality condition~(\ref{CNmom}) implies that
$\widetilde\Ihat(\zeroIII\ts) = 1$.  Furthermore, for
asymptotically large $\qIII$, $\widetilde\Ihat(\qIII\ts) \rightarrow 0$,
since in position space $\Ihat$ is reasonably smooth and 
has a finite region of support.  

Finally, the only quantity appearing in Eq.~(\ref{rhorho3}) yet to be
specified is $\Kcube$.  We determine $\Kcube$ by 
integrating the trace of~(\ref{rhorho3}) and replacing
$\DIII(\deltaIII)$ by $\delta^3(\deltaIII)$:
the result should be $3AC_F$ for a nucleus containing $A$
nucleons.  Thus
\beq
\Kcube = {{3AC_F}\over{N_c^2-1}} \ts {{1}\over{V}}
 = {{3A}\over{2N_c}} \ts {{1}\over{V}}.
\label{Kdef}
\eeq


\section{Region of Validity}\label{VALIDITY}

All of the machinery assembled in the previous two sections
has been geared towards performing a classical computation of
the vector potential associated with a color charge moving
down the $z$ axis with a speed $\beta$ near, but not equal to,
the speed of light.  The vector potential is then translated into
a gluon number density in the spirit of the
Weizs\"acker-Williams  approximation.
In this subsection we will consider the
conditions which must be satisfied in order for this
treatment to be valid.

Firstly, we need the coupling $\alpha_s = g^2/4\pi$ to be weak.
When $\alpha_s \ll 1$, we have the possibility that the
quantum corrections will be small, making the classical result
a reasonable approximation to the full result.  
Several years ago, McLerran and Venugopalan\cite{paper1} observed
that for a very large nucleus or at very small values of $\xf$
the density of quarks and gluons per unit area per unit rapidity
is large.  When this density is much larger than
$\LQCD^2$, we expect $\alpha_s$ to be 
weak\cite{paper18,paper21,paper19}.

The large density of color charge facilitates the classical
treatment in a second fashion:  when there a large number
of charges contributing to the charge density at each point,
the total will (typically) be in a large representation of the
gauge group.  Thus, we may treat the source classically.
In addition, the large number of quarks justifies the use of a
Gaussian weight for the ensemble average via the central
limit theorem.

\begin{figure}[h]

\vspace*{19cm}
\includegraphics{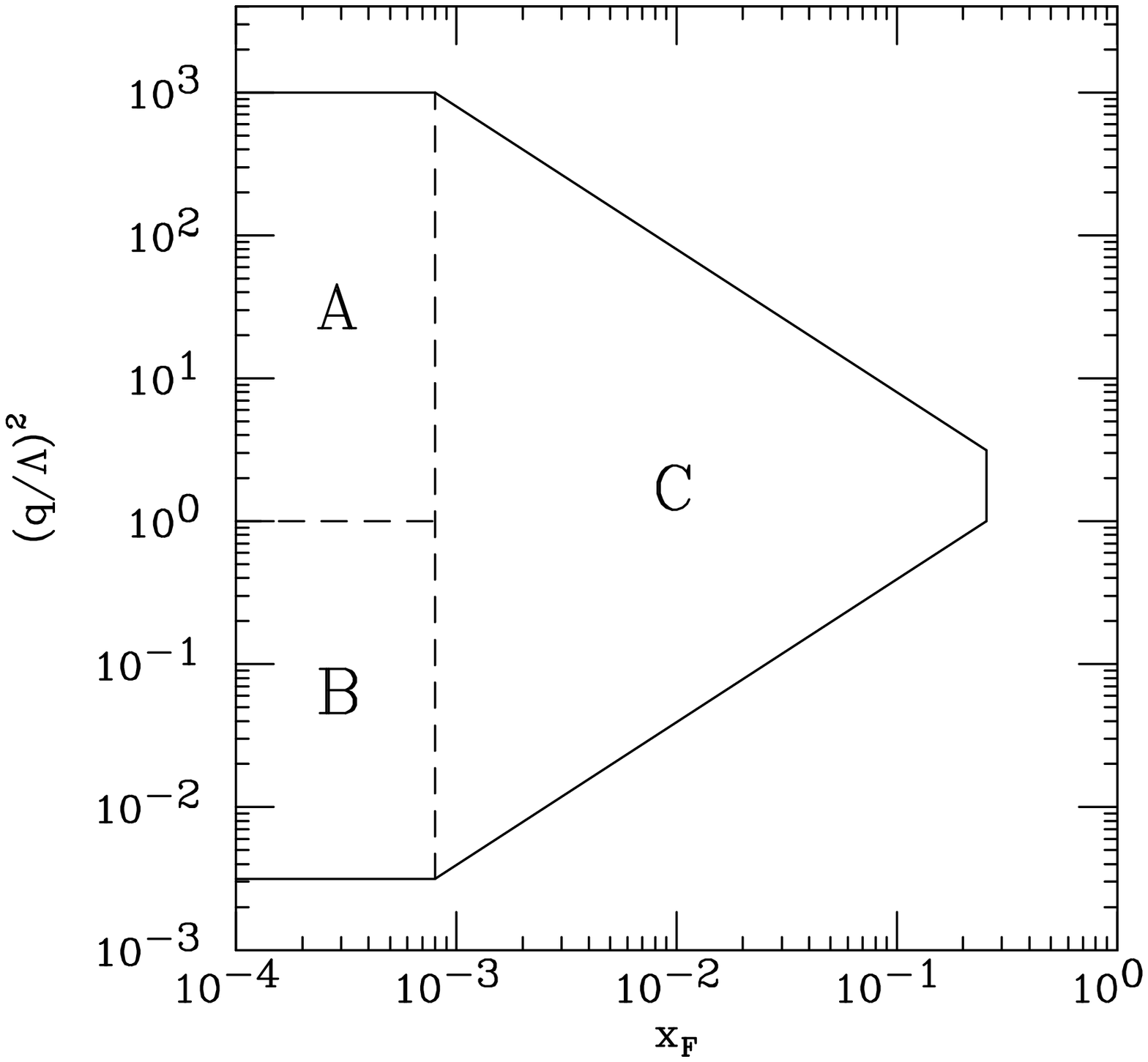}
\includegraphics{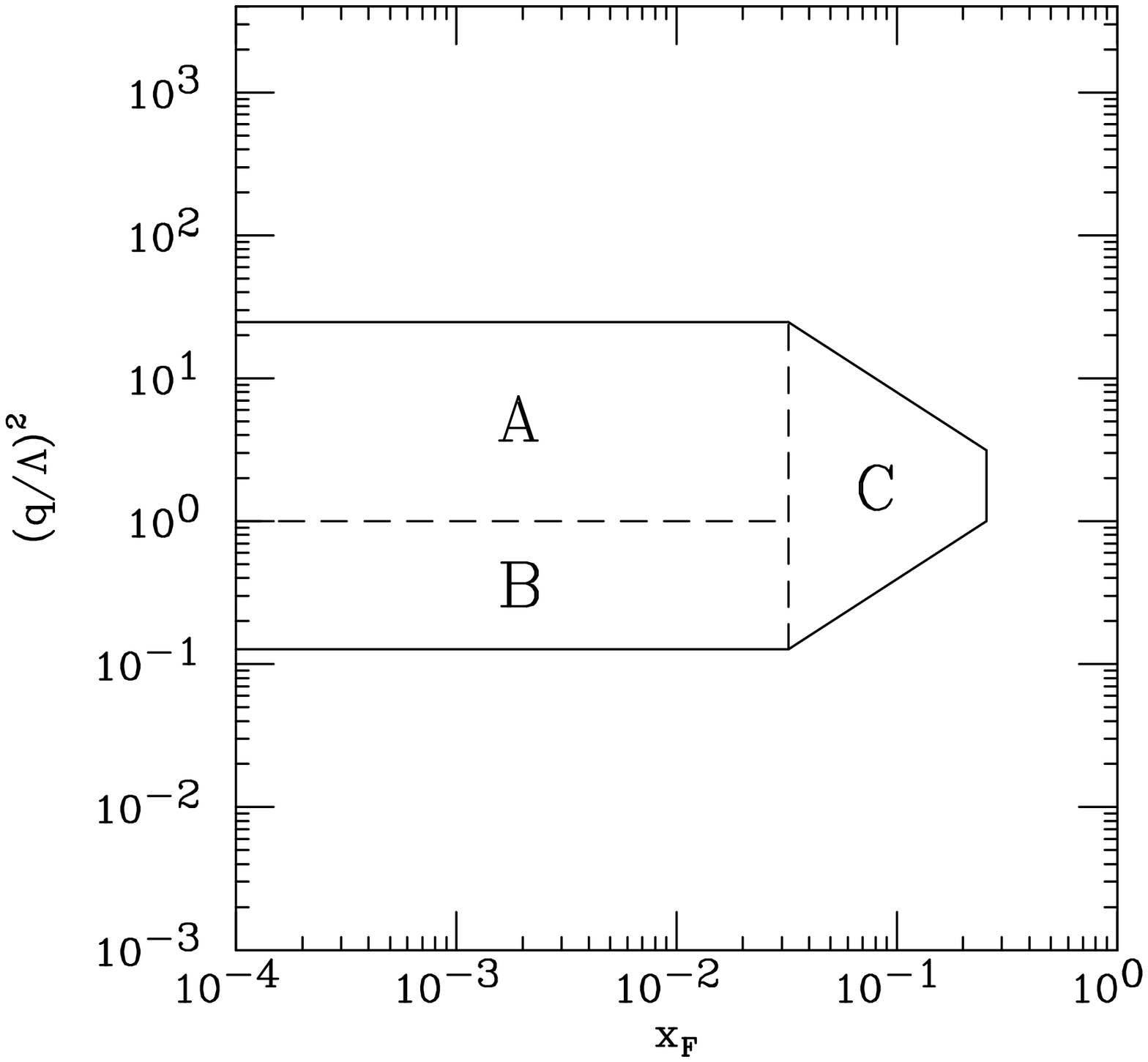}

\caption[]{Approximate region where the density of color charge
is large (and hence $\alpha_s$ weak), and sufficient color charge is
being probed to justify a classical approximation to the quantum
average in Eq.~(\protect\ref{gnum3d}).
(A) Small-$\xf$ region of the original MV 
model\protect\cite{paper1,paper2,paper3,paper4,paper9}.
(B) Additional allowed region at small $\xf$ when the
effects of confinement are included\protect\cite{PAPER14}.
(C) Extension to larger $\xf$ discussed in this paper.
The upper plot is for a toy nucleus with $A^{1/3}=250$, while
the lower plot is for uranium-238.
}
\label{ValidRegion}
\end{figure}

Under what conditions do we see a high color charge density?  
And when do a large number of charges contribute?
The answers to these questions depend upon the scale at which the
gluon distribution is being probed (see Fig.~\ref{ValidRegion}).
In the original MV treatment\cite{paper1,paper2,paper3,paper4,paper9},
it was assumed that $\xf$ was ``small.''  In this case, ``small''
means that the longitudinal scale resolved by the gluons is larger than
the Lorentz-contracted thickness of the nucleus which they see:
\beq
{{1}\over{\qz}} \agt R, \qquad\hbox{or}\qquad
\xf \alt { {A^{-1/3}}\over{\mn a} }.
\label{small-xF}
\eeq
All of the quarks at a given transverse position $\xt$
contribute to generating the gluon field measured at the values
of $\xf$ indicated in Eq.~(\ref{small-xF}).  This leads
to a color charge per unit area of
\beq
\kappa^2 \equiv
{{3AC_F}\over{\pi R^2}} \approx {{3A^{1/3}C_F}\over{\pi a^2}}.
\label{maxKAPPA}
\eeq
Asking that this density be $\agt\LQCD^2 \sim a^{-2}$ 
so that the coupling
$\alpha_s$ be weak leads to the condition
\beq
A^{1/3} \agt {{\pi}\over{3C_F}}.
\eeq
Refs.~\cite{paper2,paper3,paper4,paper9} restrict the
allowable $\qt^2$ to the region 
\beq
\LQCD^2 \alt \qt^2 \alt 3A^{1/3}C_F \LQCD^2.
\label{oldQ2limits}
\eeq
The lower limit in~(\ref{oldQ2limits}) comes from the requirement
that the gluons probe distances small compared to the nucleon radius.
On the other hand, if $\qt^2$ is too large, an insufficient number
of quarks will contribute, no matter how big the color charge density
is.  Since the amount of transverse
area probed by a gluon with transverse momentum $\qt^2$ is about
$\pi/\qt^2$, we conclude that for the charge per unit area given 
in~(\ref{maxKAPPA}), the gluon sees an amount of charge equal to
$3A^{1/3} C_F \LQCD^2/\qt^2$.  Asking that this be much greater than
one leads to the upper limit in~(\ref{oldQ2limits}).
The requirements of Eqs.~(\ref{small-xF}) and~(\ref{oldQ2limits})
restrict the range of validity of the original MV model to the
region labelled ``A'' on Fig.~\ref{ValidRegion}.

Because the strong coupling is evaluated at $\kappa^2$
rather than $\qt^2$, it ought to be possible
to relax the lower limit in~(\ref{oldQ2limits}),
provided that the theory is infrared finite.
All that is
required is a framework which captures the key consequence of
confinement, namely the fact that 
when viewed on large ($\gg a$) distance scales,
the nucleons are color neutral.  
This observation leads to the color neutrality 
condition~(\ref{CNpos}) to be imposed on the
two-point charge density correlation function\cite{PAPER14}.
Not only does the color neutrality condition make the
theory infrared finite, but it also limits the amount
of color charge being probed as $\qt\rightarrow 0$:
beyond
about $\qt^2 \sim 1/a^2$ the net charge drops as complete
color-neutral nucleons are probed. 
The net effect is to reduce $\kappa^2$ by the factor $(a\qt)^2$ from
the value given in~(\ref{maxKAPPA}), leading to the 
less-stringent lower limit
\beq
\qt^2 \agt {{\pi A^{-1/3}}\over{3C_F}} \ts  \LQCD^2.
\label{improvedQ2limit}
\eeq
The region labelled ``B'' in Fig.~\ref{ValidRegion} represents
the additional range of validity obtained in infrared finite
theories by replacing the lower limit of~(\ref{oldQ2limits})
with~(\ref{improvedQ2limit}).

Now we turn to the main goal of this paper, the relaxation
of the condition~(\ref{small-xF}) on $\xf$.
At larger values of $\xf$, the gluons are able to probe shorter
longitudinal distance scales:  they no longer see the entire
thickness of the Lorentz-contracted nucleus.  The fraction of this
thickness which  they do see is roughly 
$1/\qz R \sim 1/\mn\xf A^{1/3} a$.  
Hence, the value of $\kappa^2$ obtained in Eq.~(\ref{maxKAPPA})
is reduced by this factor, and the coupling is weak only for
\beq
\xf \alt {{3C_F}\over{\pi\mn a}}.
\label{xF-max}
\eeq
Taking $\mn a \approx 5$, Eq.~(\ref{xF-max}) implies an upper
limit of $\xf \alt 0.25$.  At larger values of $\xf$, too little
of the nucleus is seen by the gluons in order for the charge
density to be large and the coupling weak, independent of how
large we imagine the nucleus to be.

Eq.~(\ref{xF-max}) is not the final word, however.  As noted
above, when $\qt^2$ becomes small, the effective charge
density is reduced since we begin to see color neutral
nucleons.  Thus, we should further reduce $\kappa^2$ by the factor
$(a\qt)^2$ in this region, leading to the lower limit
\beq
{{\qt^2}\over{\xf}} \agt \ts {{\pi ma}\over{3C_F}} \ts \LQCD^2,
\label{q2min}
\eeq
valid whenever $\xf$ is ``large'' ({\it i.e.} when $\xf$
is larger than the value given in Eq.~(\ref{small-xF})).
Likewise, when $\qt^2$ is made too large, not enough charge
is probed.  The upper limit is reduced from the value given
in Eq.~(\ref{oldQ2limits}) by a factor of $1/\qz R$, to
\beq
\xf \qt^2 \alt {{3C_F}\over{ma}}\ts \LQCD^2.
\label{q2max}
\eeq
Taken together, the constraints~(\ref{xF-max}), 
(\ref{q2min}), and~(\ref{q2max}) allow us to extend the 
computation of the gluon number density into the region
labelled ``C'' on Fig.~\ref{ValidRegion}.

Finally, we note that the eikonal approximation
which we are using also tells us that $\qt^2$ and $\xf$
cannot get too large:  that is, we are ignoring 
nuclear recoil effects.


\section{Gluon Number Density}\label{CALCULATION}

We now turn to the computation of the gluon number density
within the 3-dimensional framework described in Secs.~\ref{SOURCE}
and~\ref{COUNTING}.
In order to obtain our result, we will have to rely on 
both color-neutrality and the large nucleus approximation
extensively.  
Our final expression reduces to the MV result of 
Refs.~\cite{paper9,PAPER14}
in the limit $\xf\rightarrow 0$, but only if the nucleus is
assumed to have cylindrical geometry.

Our calculation has three stages:  first, we obtain the solution
for the vector potential in the covariant gauge.  Next, we transform
that solution to the light-cone gauge.  Finally, we use the light-cone 
gauge solution along with the correlation function~(\ref{rhorho3})
to obtain the gluon number density from Eq.~(\ref{gnum3d}).


\subsection{Covariant Gauge Vector Potential}

The most efficient route in performing our calculation begins
by imagining the situation in the rest frame of the nucleus,
where we consider a static distribution of color charge,
Eq.~(\ref{Jrest}).  In this frame we have the ``obvious'' 
time-independent Coulomb solution for the vector potential.
Since only $A^{0}\ne 0$, we have 
$\partial_0 A^{0} = \partial \cdot A =0$,
that is, the Coulomb solution is  the same as the covariant
gauge solution.  When we boost to the lab frame then, it is 
natural to begin with the covariant gauge solution.

The Yang-Mills equations in the covariant gauge 
read\footnote{From this point forth, we will use a tilde to
distinguish the vector potential in the covariant gauge from
the vector potential in the light-cone gauge.}
\beq
(\delt^2 - 2 \partial_{+} \partial_{-}) \widetilde{A}^{\nu}
= gJ^{\nu} 
+2ig\Bl \widetilde{A}^{\mu}, \partial_{\mu} \widetilde{A}^{\nu} \Br
- ig\Bl \widetilde{A}_{\mu}, \partial^{\nu} \widetilde{A}^{\mu} \Br
+  g^2\Bl \widetilde{A}_{\mu}, \Bl \widetilde{A}^{\mu}, 
                                \widetilde{A}^{\nu} \Br\Br.
\label{x1597}
\eeq
In order to deal with these equations, we must assume not
only that the source has the form indicated in
Eq.~(\ref{Jlab}), but also that we are in the weak-coupling
regime, $g \ll 1$.  In particular, we assume that the commutator terms
appearing in~(\ref{x1597}) are negligible, leaving the simpler
equations
\beq
(\delt^2 + \dz^2 )\widetilde{A}^{\nu}(\xIII\ts)
= gJ^{\nu}(\xIII\ts).
\label{x1599}
\eeq
The operator appearing in Eq.~(\ref{x1599}) is simply the 
Laplacian in 3-dimensions.
These equations are solved in the usual manner by introducing
the Greens function  $\GIII(\xIII\ts)$ which satisfies the equation
\beq
(\delt^2 + \dz^2 )\GIII(\xIII\ts) = \delta^3(\xIII\ts).
\label{x1271-c}
\eeq
Passing to momentum space, we find that
\beq
\GIII(\qIII\ts) = 
{ {-1}\over{\qt^2 + \qz^2} }.
\label{x1272}
\eeq
The Fourier transform used to obtain~(\ref{x1272}) is easily
inverted, producing
\beq
\GIII(\xIII\ts) = 
-{ {1}\over{4\pi} } \ts
{ {1}\over\sqrt{\xt^2+\xz^2} }.
\label{x1594}
\eeq
Actually, 
because of the unequal treatment of the longitudinal and transverse
variables when we transform to the light-cone gauge,
the following mixed representation,
\beq
\GIII(\xIII\ts) =
- \int { {d^2\qt}\over{4\pi^2} } \ts
{{1}\over{2q}}\ts
e^{-i\qt\cdot\xt} \ts e^{-q\vert\xz\vert},
\label{x1400}
\eeq
which is obtained by inverting only the longitudinal part of the
transform, will prove to be especially useful.
Note that in Eq.~(\ref{x1400}), as elsewhere in this
paper, $q$ means $\vert\qt\vert$.

Independent of how we choose to write down the Greens function,
the solution to~(\ref{x1599}) with the source~(\ref{Jlab}) reads
\beqa
\widetilde{A}^{+}(\xIII\ts) &=&
{{1}\over{\EPS}}\ts g
\int d^3\xIII\ts' \ts
\GIII(\xIII-\xIII\ts') \rho(\xprimeIII)
\cr
\widetilde{A}^{-}(\xIII\ts)
 &=& \half\EPS^2 \widetilde{A}^{+}(\xIII\ts),
\phantom{\biggl[}
\cr
\widetilde{A}^j(\xIII\ts) &=& 0.
\label{x1599-b}
\eeqa


\subsection{Transformation to the Light Cone Gauge}

At this stage, we are ready to perform the transformation
to the light-cone gauge.  In a non-Abelian theory, we may
parameterize the gauge transformation as
\beq
A^{\mu}(x) = \U(x) \widetilde{A}^{\mu}(x) \U^{-1}(x)
           -{{i}\over{g}}\ts\Bigl[ \partial^{\mu}\U(x) \Bigr]\U^{-1}(x).
\label{QCDGaugeTrans}
\eeq
Since the potentials we are dealing with are functions of
$\xIII$ only, we expect that $\U$ will also
depend on $\xIII$ only.  Thus, the requirement that the
new gauge be the light-cone gauge becomes
\beq
\dz\U(\xIII\ts)
= ig\EPS\U(\xIII\ts)\widetilde{A}^{+}(\xIII\ts),
\label{Ueqn}
\eeq
where we have replaced $\partial_{-}$ by $\dz$
in accordance with~(\ref{x1594-x}).
The solution to Eq.~(\ref{Ueqn}) is the path-ordered exponential
\beqa
\U(\xIII\ts) & \equiv & 
\overline\pexp\Biggl[ ig \int_{-\infty}^{\xz} \nts d\yz
              \ts\EPS\widetilde{A}^{+}(\yz,\xt) \Biggr]
\cr \phantom{\Biggl[} &=&
\openone +
\sum_{m=1}^{\infty} \ts (ig)^m
\int_{-\infty}^{\xz} d^m{\yz}_{\downarrow} \ts\ts
\EPS\widetilde{A}^{+}({\yz}_m,\xt)\cdots
\EPS\widetilde{A}^{+}({\yz}_2,\xt)
\EPS\widetilde{A}^{+}({\yz}_1,\xt).
\label{Udef}
\eeqa
In Eq.~(\ref{Udef}) we have introduced the shorthand notation
\beq
\int_{-\infty}^{\xz} d^m{\yz}_{\downarrow}
\equiv
\int_{-\infty}^{\xz} d{\yz}_1
\int_{-\infty}^{{\yz}_1} d{\yz}_2
\cdots
\int_{-\infty}^{{\yz}_{m-1}} d{\yz}_m
\eeq 
to indicate the ordered integration region 
$\xz\ge{\yz}_1\ge{\yz}_2\ge \cdots \ge{\yz}_m > -\infty$.
Eq.~(\ref{x1599-b}) tells us that
$\widetilde{A}^{+}$ is naturally of order $1/\EPS$:
therefore,
all of the terms in the sum on the right hand side of~(\ref{Udef})
are of order unity.
Introducing the expression for $\widetilde{A}^{+}$ 
into the expression for $\U(\xIII\ts)$ produces
\beq
\U(\xIII\ts) =
\overline\pexp\Biggl[\ts ig^2 \nts\int_{-\infty}^{\xz} \nts d\yz
              \int d^3\xiIII\ts
              \ts\GIII(\yz-\xiz,\xt-\xit)\rho(\xiIII\ts) \Biggr].
\label{x1606-a}
\eeq
Inserting this into Eq.~(\ref{QCDGaugeTrans}), we find that
the transverse components of the vector potential read\footnote{The 
longitudinal component,
$A^{-}(\xIII\ts)$ turns out to be of order $\EPS$.  In any event,
it does not contribute to the gluon number density, as it does not
represent a physical gluonic polarization state.}
\beqa
A^{j}(\xIII\ts) &=& g
\int_{-\infty}^{\xz} d\yz \ts\ts
\U(\yz,\xt)
\biggl[ \int d^3\xiIII\ts\ts
 \partial^j \GIII(\yz-\xiz,\xt-\xit)\rho(\xiIII\ts) \biggr]
\Uinv(\yz,\xt) 
\cr &=& g
\sum_{m=1}^{\infty}
(-ig^2)^{m-1}  
\int_{-\infty}^{\xz} d^m{\yz}_{\downarrow} 
\int d^3\xiIII_1 \ts
\partial^j\GIII({\yz}_1-{\xiz}_1;\xt-\xit_1)
\cr  && \qquad\qquad\times
\Biggl(\prod_{\ell=2}^{m}\int d^3\xiIII_\ell \ts
\GIII({\yz}_\ell-{\xiz}_\ell;\xt-\xit_\ell)\Biggr)
\Lbrk
\rho(\xiIII_1)
\rho(\xiIII_2) \cdots 
\rho(\xiIII_m)
\Rbrk.
\label{x1606-b}
\eeqa
The quantity in the double square brackets
appearing in the last line of Eq.~(\ref{x1606-b}) is simply
a multiple nested commutator:
\beq
\Lbrk
\rho(\xiIII_1)
\rho(\xiIII_2)  \cdots 
\rho(\xiIII_m)
\Rbrk
\equiv 
\Bl\Bl\Bl \cdots \Bl  \rho(\xiIII_1),
                      \rho(\xiIII_2)\Br,
                      \rho(\xiIII_3)\Br,
          \cdots \Br, \rho(\xiIII_m)\Br.
\eeq
At this stage there are no more explicit factors of $\EPS$
appearing in the vector potential or in the expression
for the gluon number density~(\ref{gnum3d}).  Thus, the
$\EPS\rightarrow 0$ limit is trivial to perform.

The expression for the gluon number density involves 
the partially Fourier-transformed quantity $A(\qz;\xt)$.  Since
the only dependence on $\xz$ itself appearing in 
Eq.~(\ref{x1606-b}) is as the upper limit of the outermost
of the ordered integrations, we have
\beqa
A^{j}(\qz;\xt) = 
{{g}\over{i\qz}}
\sum_{m=1}^{\infty}
(-ig)^{m-1}  &&
\int_{-\infty}^{\infty}  d^m{\yz}_{\downarrow} 
\ts \exp(-i\qz{\yz}_1)
\int d^3\xiIII_1 \ts
\partial^j\GIII({\yz}_1-{\xiz}_1;\xt-\xit_1)
\cr  \times &&
\Biggl(\ts\prod_{\ell=2}^{m}\int d^3\xiIII_\ell \ts
\GIII({\yz}_\ell-{\xiz}_\ell;\xt-\xit_\ell)\Biggr)
\Lbrk
\rho(\xiIII_1)
\rho(\xiIII_2) \cdots 
\rho(\xiIII_m)
\Rbrk.
\label{x1608-a}
\eeqa
Finally, we insert the mixed representation of the Greens function 
presented in Eq.~(\ref{x1400}).  The transverse part
of the resulting $\xi$ integrations simply Fourier transforms
the transverse part of the charge densities:
\beq
\rho(\xiz; \pt) \equiv
\int d^2\xit \ts e^{i\pt \cdot \xit } \ts 
\rho(\xiIII\ts).
\eeq
Hence, the light-cone gauge vector potential becomes
\beqa
A^{j}(\qz;\xt) = 
{{1}\over{ig\qz}} 
\sum_{m=1}^{\infty} 
(ig^2)^{m}   &&
\int_{-\infty}^{\infty} d^m{\yz}_{\downarrow} 
\int_{-\infty}^{\infty} d^m\xiz
\ts \exp(-i\qz{\yz}_1)
\cr \times &&
\int {{d^2\pt_1}\over{4\pi^2}} \ts
{{p_{1j}}\over{2p_1}} \ts
e^{-i\pt_1\cdot\xt} 
\exp\Bigl(-p_1\vert{\yz}_1-{\xiz}_1\vert\Bigr)
\cr  \times &&
\prod_{\ell=2}^{m}
\int {{d^2\pt_\ell}\over{4\pi^2}} \ts
{{1}\over{2p_\ell}} \ts
e^{-i\pt_\ell\cdot\xt} 
\exp\Bigl(-p_\ell\vert{\yz}_\ell-{\xiz}_\ell\vert\Bigr)
\cr \times &&
\Lbrk
\rho({\xiz}_1;\pt_1)
\rho({\xiz}_2;\pt_2) \cdots 
\rho({\xiz}_m;\pt_m)
\Rbrk. \phantom{\biggl[}
\label{x1609}
\eeqa

\begin{figure}[h]

\vspace*{17cm}
\includegraphics{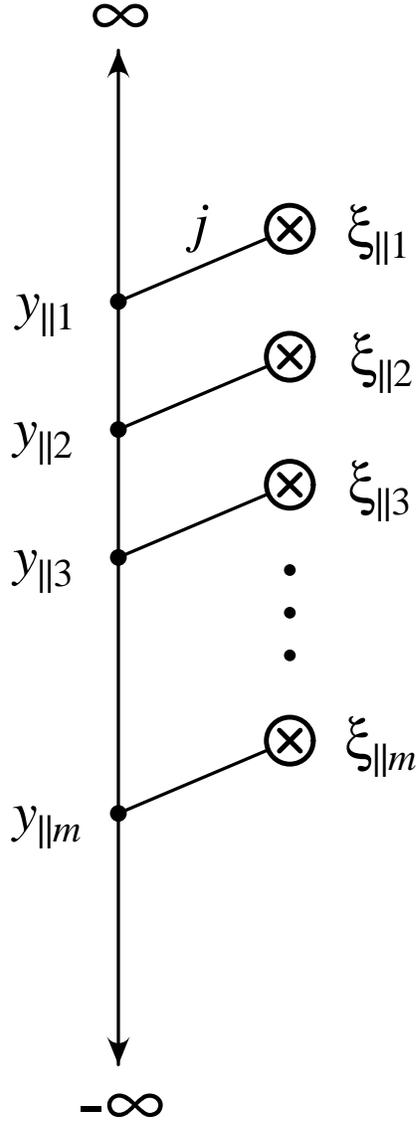}

\caption[]{Diagrammatic representation of longitudinal
structure of the $m$th-order term
in the expression for the light-cone gauge vector potential,
Eq.~(\protect\ref{x1609}).  The circled crosses denote the
positions at which the sources are being evaluated.  The dots
represent the ordered integrations coning from the gauge transformation
into the light-cone gauge.
}
\label{DiagA1}
\end{figure}

It is useful to have a diagrammatic representation of the
contributions to Eq.~(\ref{x1609}).  
Because it turns out that the longitudinal structure is
significantly more complicated than the transverse structure,
our diagrams
are meant as an aid in understanding the 
longitudinal structure.
Fig.~\ref{DiagA1} illustrates the $m$th term of Eq.~(\ref{x1609}).
The vertical line represents the range of the $\yz$ integrals,
with each of the vertices (dots) being the value of one of the 
$\yz$'s.  Because these integrations are ordered, the dots
are not allowed to slide past each other.  The sources, whose 
longitudinal coordinates are $({\xiz}_1,{\xiz}_2,\ldots,{\xiz}_m)$,
are denoted by the circled crosses.  The lines connecting
the $\yz$'s with the $\xiz$'s correspond to the longitudinal
factors of the Greens functions, 
$\exp(-p_i\vert{\yz}_i-{\xiz}_i\vert)$.
Finally, the index $j$ labelling the ${\yz}_1$-${\xiz}_1$ line 
reminds us that it is special:  not only is 
the Greens function associated with this factor differentiated
(producing the factor of $p_{1j}$), but there is a factor 
$\exp(-i\qz{\yz}_1)$ left over from the longitudinal Fourier transform
which was performed on the vector potential.


\subsection{Determining the Gluon Number Density}

We now turn to the computation of the gluon number density.
Because of the extended longitudinal structure, the calculation
is rather lengthy.  Here we will outline the path to the result
with the help of our diagrammatic representation.  The mathematical
details appear in Appendix~\ref{CONTRACTIONS}.  

The full calculation of the gluon number density essentially consists
of inserting two copies of Eq.~(\ref{x1609}) for the vector
potential into the master formula~(\ref{gnum3d}) for
the gluon number density, and performing all possible pairwise
contractions of the sources in each term of the result.
We retain
only the leading terms in powers of $a/R$.
There are a total of four longitudinal integrations
per contracted pair of sources:  the (unordered) position $\xiz$ 
associated with the inversion of the Yang-Mills equation~(\ref{x1599})
for each source  (represented by the circled crosses in Fig.~\ref{DiagA1})
plus the (ordered) integration on $\yz$ associated
with the transformation to the light-cone gauge (represented by the
points on the vertical line).  A priori,
these integrations could produce a factor of $R^4$ (per pair).  
However, we shall now argue that at most they produce a factor 
of $a^3 R$.  The key observations to make are that both the propagators
and the correlation function $\langle\rho\rho\rangle$ (through $\DIII$)
allow for a longitudinal separation 
between the points they connect which is at most of order $a$.
For the $\langle\rho\rho\rangle$ correlator, this property 
follows immediately from the color neutrality condition:  the
confining nature of QCD tells us that $\DIII(\xIII-\xprimeIII)$
should be negligible when the two points being compared are
separated by more than the nucleon size $a$ (transversely or
longitudinally, since we take $\DIII$ to be spherically
symmetric).  
In the
case of the propagator, the mixed form~(\ref{x1400}) is particularly
illuminating:  the longitudinal separation of the two points
must be order $1/p$ or less (where $p$ is the typical transverse
momentum flowing in the propagator) to avoid exponential damping
of the contribution.  However, the $\langle\rho\rho\rangle$
correlation function limits the transverse momenta to the region
$ p \agt 1/a$, again because of color neutrality.
Since the four points in question are connected via two propagators
and one contraction, three of the four integrations are restricted
to have range $a$, while the remaining integration has the
potential to roam freely over the full range of order $R$.

\begin{figure}[h]

\vspace*{20cm}
\includegraphics{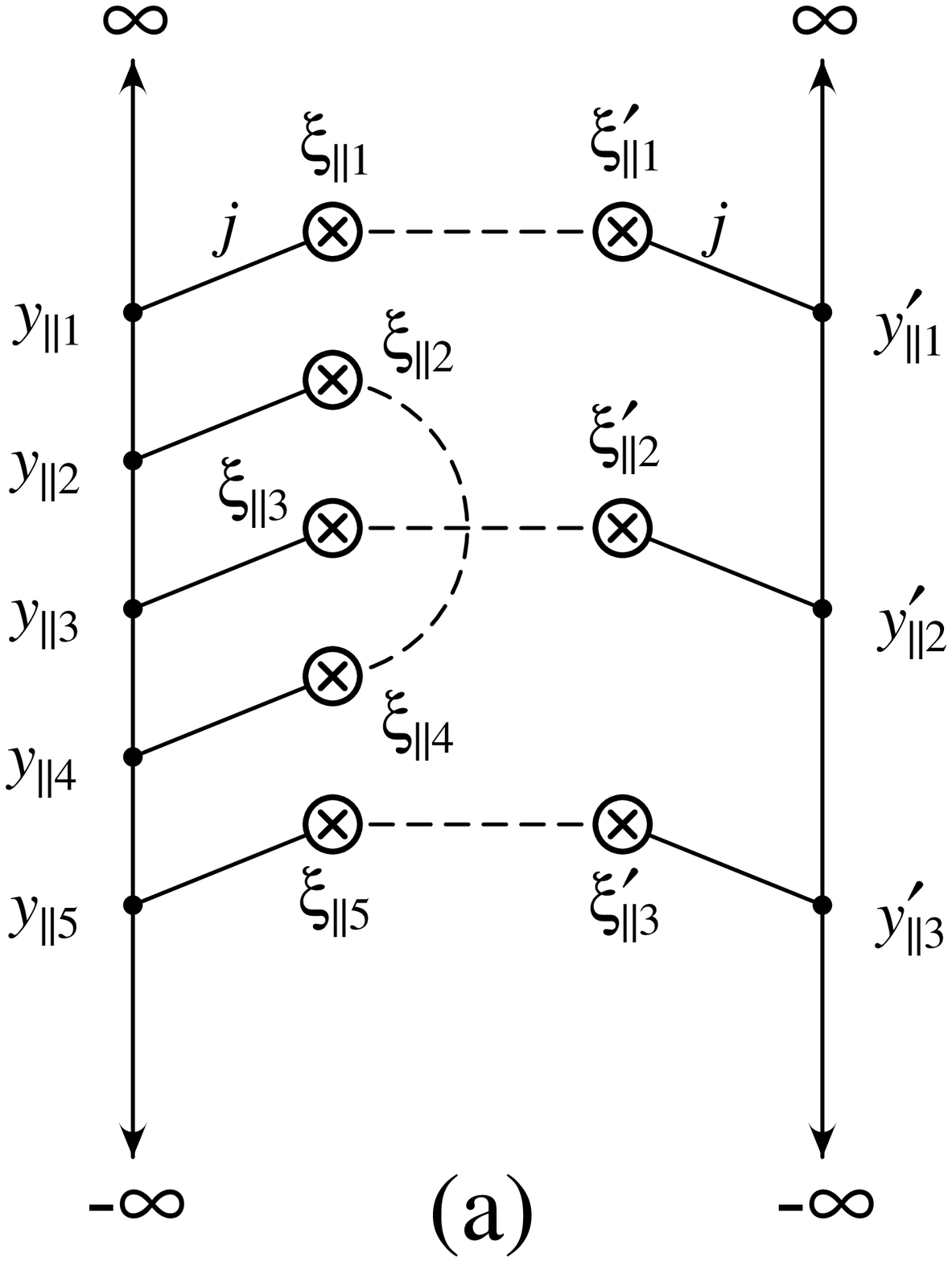}
\includegraphics{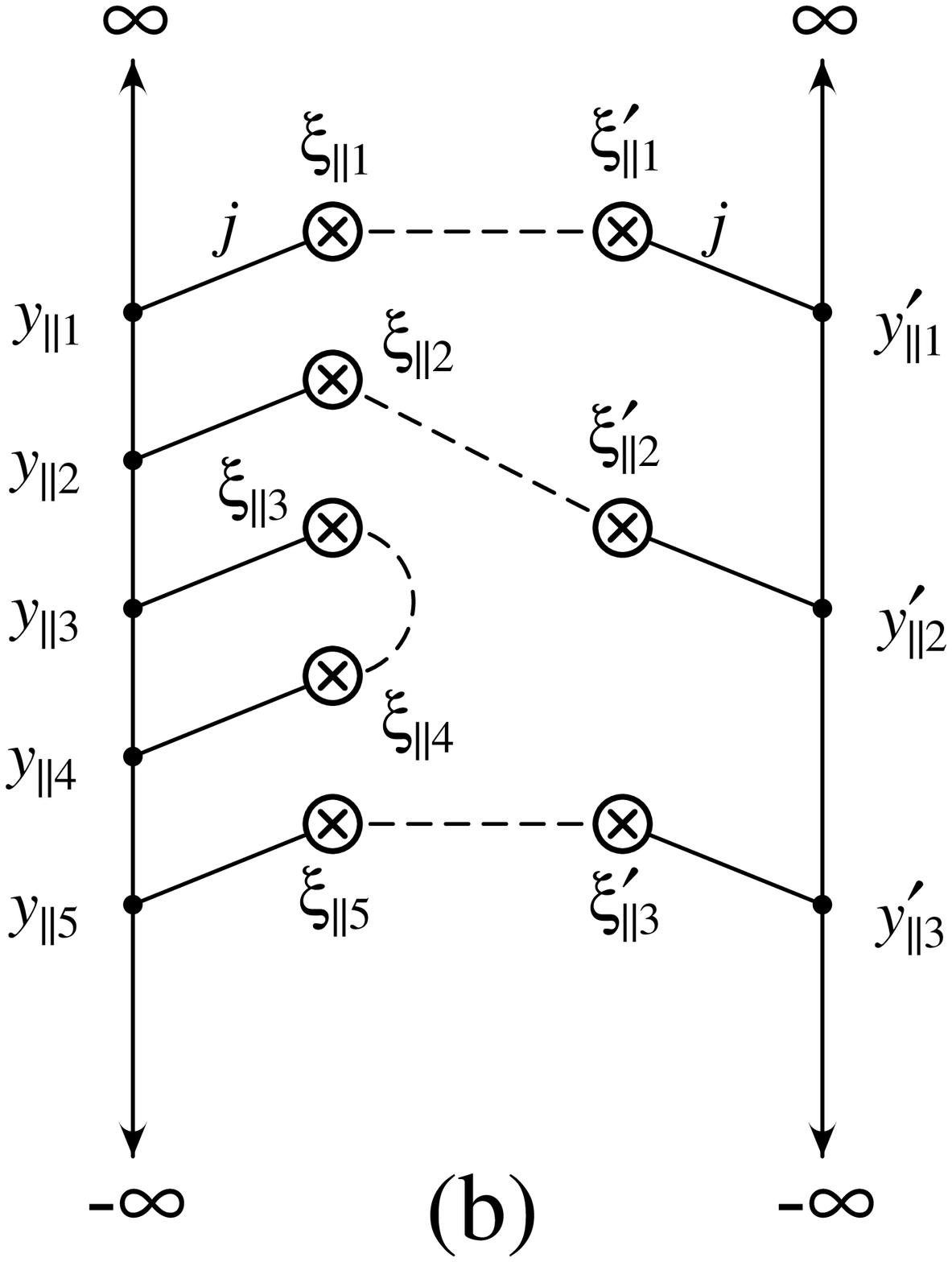}
\includegraphics{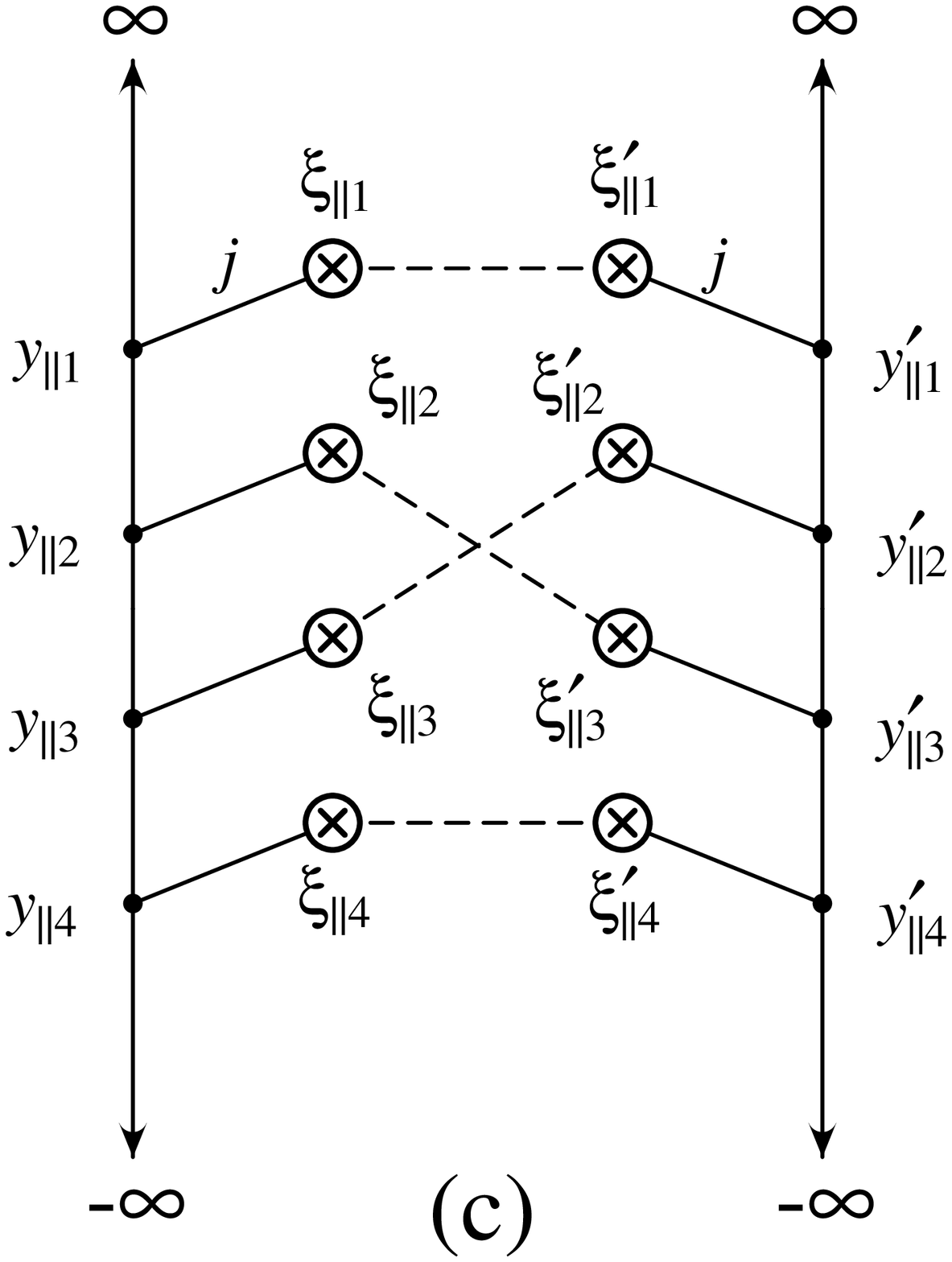}
\includegraphics{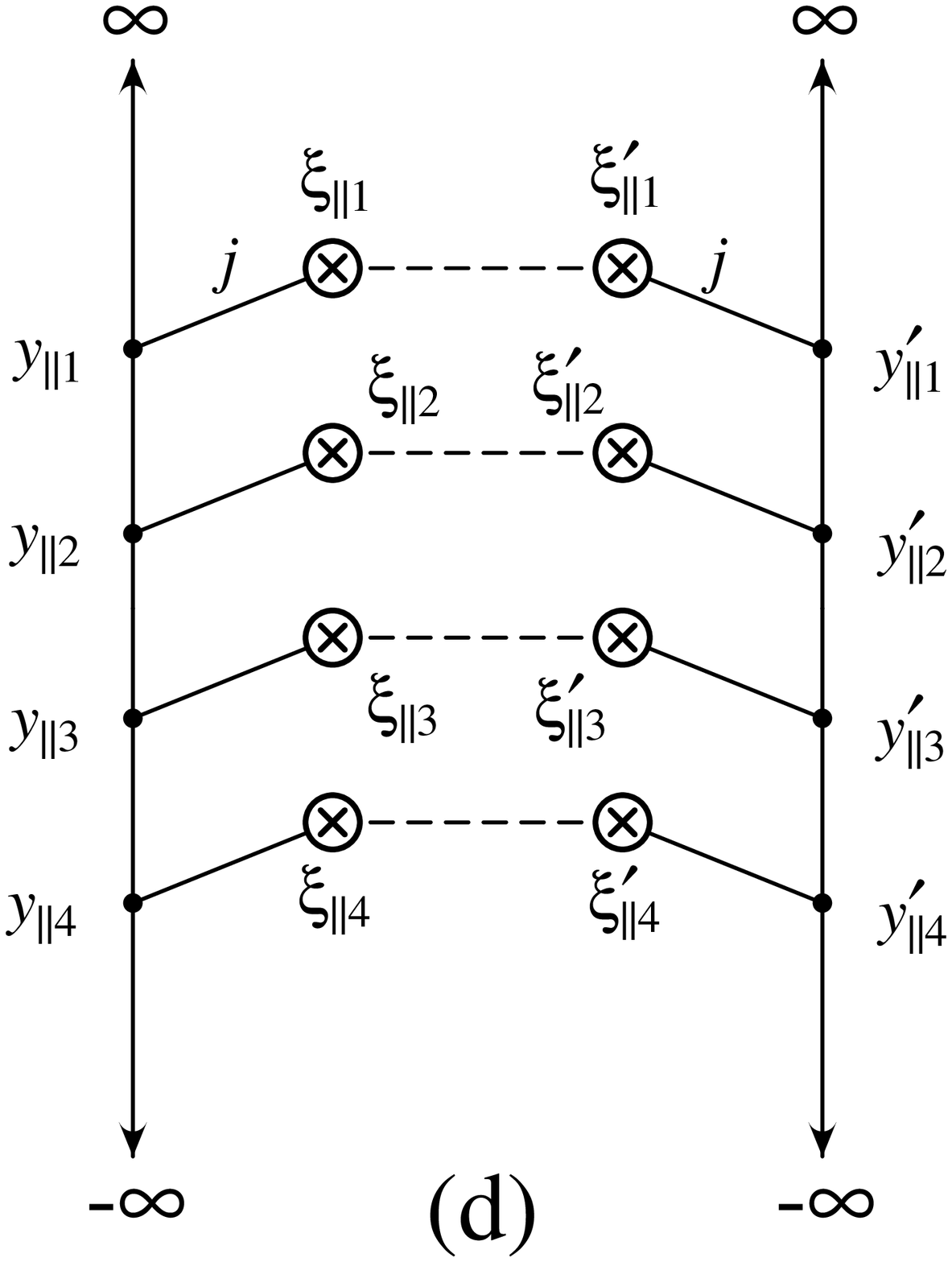}

\caption[]{Some of the contributions to the gluon number density
at 8th order in $\rho$.  (a) Diagram containing a non-adjacent
self-contraction.  (b)  Diagram containing an adjacent self-contraction.
(c) Diagram containing a pair of crossed mutual contractions.
(d) Diagram containing only corresponding mutual contractions.
Diagrams (a) and (c) are suppressed by a power of $a/R$ relative
to diagrams (b) and (d).
}
\label{EighthOrder}
\end{figure}

Nevertheless, not all combinations of contractions produces the maximum
factor $a^3 R$ for all pairs.  Fig.~\ref{EighthOrder} illustrates
some of the possible contributions at 8th order in $\rho$.
From the previous paragraph we know that
each of these diagrams could, at most, contribute four powers of $R$.
However,
the contribution in Fig.~\ref{EighthOrder}a contains only three
powers of $R$:  the self-contraction connecting ${\xiz}_2$
with ${\xiz}_4$ effectively forces ${\yz}_2$ and ${\yz}_4$ to
be at most a distance $a$ apart.  But this hems in the point
at ${\yz}_3$, preventing it from independently spanning the 
full range $R$.
On the other hand, the self-contraction in Fig.~\ref{EighthOrder}b
does not restrict the range of any additional ${\yz}$'s:  it
contains the maximum four powers of $R$.
Likewise, the set of mutual contractions illustrated in
Fig.~\ref{EighthOrder}c produce only three powers of $R$, since the
two ``crossed'' contractions cannot slide up and down independently.
In contrast, Fig.~\ref{EighthOrder}d contributes the full four
powers of $R$, since all of the ``rungs'' may move freely through
the full vertical range.

Thus, the computation in Appendix~\ref{CONTRACTIONS} includes
all diagrams which contain only uncrossed mutual contractions
(like Fig.~\ref{EighthOrder}d) or any combination of self-contractions
between adjacent sources plus uncrossed mutual contractions
(like Fig.~\ref{EighthOrder}c).  This produces the leading 
behaviour in the limit $a/R \ll 1$.
The final result reads
\beqa
\gnumX = 
3AC_F \ts
{ { 2\alpha_s } \over{\pi^2} } \ts 
{ {1}\over{\xf} }  &&
\int d^2\deltat \ts
e^{i\qt\cdot\deltat} \curlyL(\xf;\deltat) \ts
\curlyE\Bigl( \vsqr L(\deltat) \Bigr),
\label{x1629}
\eeqa
where
\beq
\curlyL(\xf;\deltat) \equiv {{1}\over{2}}
\int { {d^2\pt}\over{4\pi^2} } \ts
e^{-i\pt\cdot\deltat} \ts
{
{ \pt^2 \ts \widetilde{\DIII}(\xf\mn,\pt) }
\over
{ \Bigl[ \pt^2 + (\xf\mn)^2 \Bigr]^2 }
},
\label{x1490}
\eeq
and
\beq
L(\deltat)   \equiv
\int { {d^2\pt}\over{4\pi^2} } \ts
{
{ \widetilde{\DIII}(0,\pt) }
\over
{  \pt^4  }
}
\ts \Bigl[
e^{-i\pt\cdot\deltat}-1
\Bigr] .
\label{x1488a}
\eeq
Despite their superficial appearance,
these functions are infrared finite
for a spherically symmetric correlation function which
satisfies the color neutrality condition~(\ref{CNmom}).
In the Abelian limit ($\alpha_s^2 A^{1/3}\rightarrow 0$),
the gluon number density is simply 
\beqa
\gnumX\Biggl\vert_{{\rm lowest}\atop{\rm order}} \nts\nts &=&
3AC_F \ts
{ { 2\alpha_s } \over{\pi^2} } \ts 
{ {1}\over{\xf} }  
\widetilde\curlyL(\xf;\qt).
\cr &=&
3AC_F \ts
{ { \alpha_s } \over{\pi^2} } \ts 
{ {1}\over{\xf} }  
{
{ \qt^2 \ts \widetilde{\DIII}(\xf\mn,\qt) }
\over
{ \Bigl[ \qt^2 + (\xf\mn)^2 \Bigr]^2 }
}.
\label{AbelianLimit}
\eeqa
The nuclear correction function $\curlyE$ encodes 
the non-Abelian effects and depends on the geometry 
and size of the nucleus:
\beq
\curlyE(z) = 
\cases{ \displaystyle{{{1}\over{z}}\ts ( e^z - 1),}
                     & (cylindrical);\cr
\phantom{x} & \cr
\displaystyle{{{3}\over{z^3}} \ts
\Bigl[  1 - \half z^2 + e^z(z-1) \Bigr],  }
                     & (spherical).\cr
}
\label{x1629U}
\eeq
In either case, we have
\beq
\lim_{z\rightarrow0} \ts\curlyE(z) = 1.
\label{Elim}
\eeq
Finally, the magnitude of $\vsqr$
governs the relative importance of the nuclear corrections:
\beq
\vsqr =
\cases{ \displaystyle{{3A g^4}\over{2\pi R^2}} \approx
                      24\pi\alpha_s^2 A^{1/3} \LQCD^2 , 
                       & (cylindrical);\cr
\phantom{x} & \cr
\displaystyle{{9A g^4}\over{4\pi R^2}} \approx
                      36\pi\alpha_s^2 A^{1/3} \LQCD^2 , 
                       & (spherical).\cr}
\label{x1629V}
\eeq
These corrections are enhanced for very large nuclei.
In writing down Eqs.~(\ref{x1629U})--(\ref{x1629V}) we
have assumed that the nucleons are uniformly distributed
within the volume of the nucleus.

The result presented in Refs.~\cite{paper9,PAPER14} is recovered 
in the $\xf\rightarrow 0$ limit by
using {\it cylindrical}\ geometry, since these papers
assume that $\mu^2$
({\it i.e.}\ the part of the correlation function which corresponds
to $\SIII$ in the present paper) is a function of the longitudinal
coordinate only.  This is only true for a cylindrical nucleus.
Actually, the difference between the two functions in 
Eq.~(\ref{x1629U}) is very small when the different values of
$\vsqr$ indicated in Eq.~(\ref{x1629V}) are taken into account.

Away from $\xf= 0$ there are two distinct sources
of finite $\xf$ corrections:  the correlation function 
$\widetilde\DIII(\xf\mn,\pt)$, and the propagator appearing
in the Abelian result.
What is perhaps surprising is the fact that
the function $L(\deltat)$ turns out to be identical to 
its 2-D counterpart: it depends only on the value
of $\widetilde{\DIII}$ at $\xf=0$.


\subsection{General Properties of the Gluon Number Density}\label{GENERAL}

Because of the similarity of Eqs.~(\ref{x1629})--(\ref{x1629V})
to the result obtained in Ref.~\cite{PAPER14}, the properties
of the gluon number density which were described in Sec.~IV
of that paper continue to hold.  
In particular, 
the $\deltat\rightarrow 0$ behaviour of Eq.~(\ref{x1490})
is unchanged from the behaviour of 
its counterpart in Ref.~\cite{PAPER14}.
Thus,
we still have the transverse momentum sum rule
\beq
\int d^2\qt \ts
\Biggl\{ 
\gnumX\Biggl\vert_{{\rm all}\atop{\rm orders}} \nts\nts
-  \ts \gnumX\Biggl\vert_{{\rm lowest}\atop{\rm order}} 
\Biggr\}
= 0,
\label{SumRule}
\eeq
even when $\xf\ne 0$.
Eq.~(\ref{SumRule}) states that the nuclear corrections
have no effect on the total number of gluons at each value of $\xf$:
we could have
obtained the same number of gluons by ignoring the non-linear
terms in the light-cone gauge vector potential.  
What these corrections
actually do is to move gluons from one value of the transverse
momentum to another.  Thus, the total energy in the gluon field
at a given value of $\xf$ 
is affected by the non-Abelian terms.
The gluon structure function resolved at the scale $Q^2$
is obtained by integrating the fully differential number density
over transverse momenta satisfying $\vert\qt\vert \le Q$
(see Eq.~(\ref{StrFnDef})).  Consequently, the transverse momentum
sum rule tells us that for large values of $Q^2$, the non-Abelian
effects die off, reflecting the expected asymptotic freedom
of the theory.
We should caution, however, that unless $\xf$ is very small,
the maximum $Q^2$ for which our treatment is valid is not very
large (see Fig.~\ref{ValidRegion}).  Thus, we conclude that at 
such values of $\xf$ the non-Abelian terms are always important 
at the (smallish) values of $Q^2$ for which our approximations
hold.

\begin{figure}[h]

\vspace*{16cm}
\includegraphics{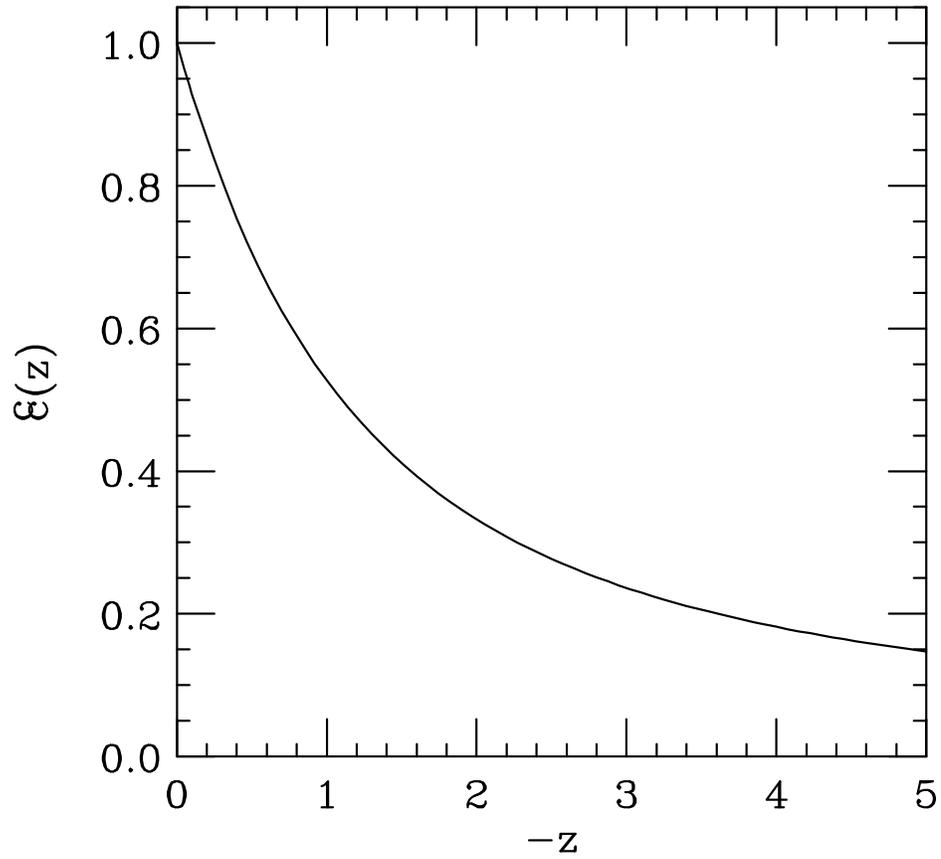}

\caption[]{Plot of the nuclear correction function
$\curlyE(z)$ for negative values of $z$
and a spherical nucleus.
}
\label{Efig}
\end{figure}

\begin{figure}[h]

\vspace*{15cm}
\includegraphics{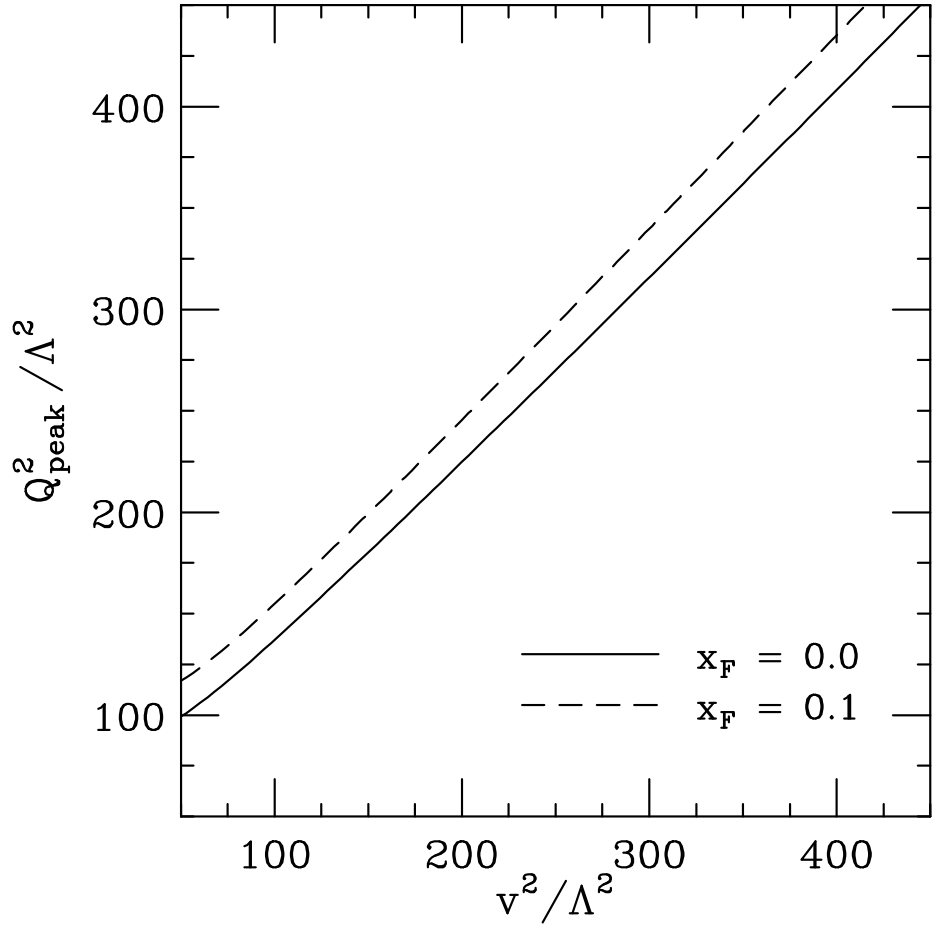}

\caption[]{Value of the momentum-squared at the peak of the
$\qt^2 dN/d\xf d^2\qt$ distribution as a function of
the scale set by the non-Abelian corrections,
$v^2 \propto A^{1/3}\LQCD^2$.  These curves have been generated
within the $\omega=1$ power-law model.
}
\label{PeakPlot}
\end{figure}

The over-all shape of the fully differential gluon number distribution
is insensitive to the detailed nucleon structure incorporated
in $\DIII$.  Instead, it is fixed only by the confinement scale
$a \sim \LQCD^{-1}$ 
plus the relative 
importance of the nuclear corrections, governed by $\vsqr$.  
Recall that according to the discussion of
Eqs.~(\ref{x1629})--(\ref{x1629V}), the nuclear corrections
are contained in the function $\curlyE(v^2L(\deltat))$.
To understand how $\curlyE$ behaves, we need to know two facts
about $L(\deltat)$.  Firstly, according to Eq.~(\ref{x1488a}),
$L(\zero) = 0$.  Therefore, at large $\qt^2$, the all-orders
distribution is identical to the lowest-order result.
(This is one of the observations used in
Ref.~\cite{PAPER14} to derive the transverse momentum
sum rule.)
Secondly, for a wide range of physically reasonable
choices for $\DIII(\xIII-\xprimeIII)$, 
$L(\deltat)\le 0$, and decreases as $\vert\deltat\vert$
increases\cite{PAPER14}.
Thus, we are interested in the behaviour of $\curlyE$ for negative 
values of its argument, which we have displayed in Fig.~\ref{Efig}.
From the figure it is easy to see that the long-distance 
contributions to the integrand of~(\ref{x1629}) are damped by
the presence of $\curlyE$.  
This behaviour is consistent with the confining nature of QCD, 
which we incorporated into the form chosen for $\DIII$
by requiring it to obey the color-neutrality
condition.   Independent of the other details of $\DIII$,
we find that 
\beq
\lim_{\qt\rightarrow 0} \ts\ts
\xf \gnumX\Biggl\vert_{{\rm all}\atop{\rm orders}} \nts = \ts
constant,
\label{qzero}
\eeq
that is, the distribution saturates as $\qt^2$ is 
lowered.\footnote{The lowest order result~(\protect\ref{AbelianLimit})
actually vanishes at $\qt^2=0$ when $\xf=0$.  This may be viewed
as an accidental cancellation in the integrand of~(\ref{x1629})
when $\curlyE\equiv1$.}  Furthermore, since increasing the size
of the nucleus increases $\vsqr$, which in turn increases the
amount of damping provided by $\curlyE$ for the same value of $\deltat$,
the gluon density at which saturation occurs decreases as
$A^{1/3}$ is increased.
Because the large-$\qt$ part of the distribution does
not change,
we conclude that in order to satisfy the sum rule~(\ref{SumRule}),
the number of gluons at intermediate momenta must increase.
Heuristically, the position space width of the non-Abelian factor 
$\curlyE(\vsqr L(\deltat))$ goes like $v^{-1}$.
This width provides a second length scale in addition to 
the scale $\LQCD^{-1}$ characteristic of the lowest-order result.
Thus, we might expect that momenta of order $\qt^2 \sim \vsqr$
would play an important role in the resulting all-orders distribution.
According to Eq.~(\ref{x1629V}), we expect $\vsqr\propto A^{1/3}\LQCD^2$.
This is, in fact, what we observe in our numerical calculations:
an enhancement in the number of gluons with transverse momenta
of order $\vsqr$
(see Fig.~\ref{PeakPlot} in Sec.~\ref{EXAMPLE}).
Although the idea that a new scale proportional to $A^{1/3}\LQCD^2$
should emerge and play an important role for large enough nuclei is not 
new\cite{paper1,paper2,paper9,PAPER14,paper18,paper21,paper19,%
paper37,paper34,LatSat}
our results lend further support to this concept.


\section{Illustration of Our Results}\label{EXAMPLE}


\subsection{Power-Law Model for $\widetilde\DIII(\qIII\ts)$}

\begin{figure}[h]

\vspace*{20cm}
\includegraphics{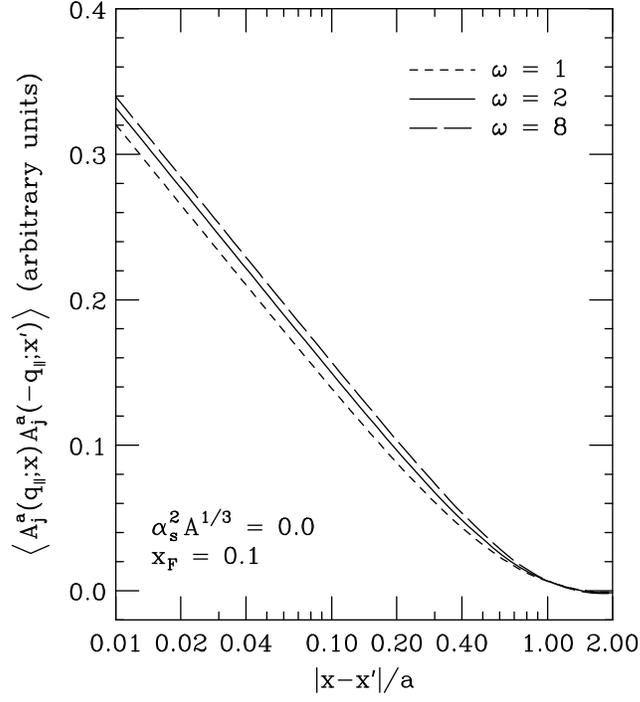}
\includegraphics{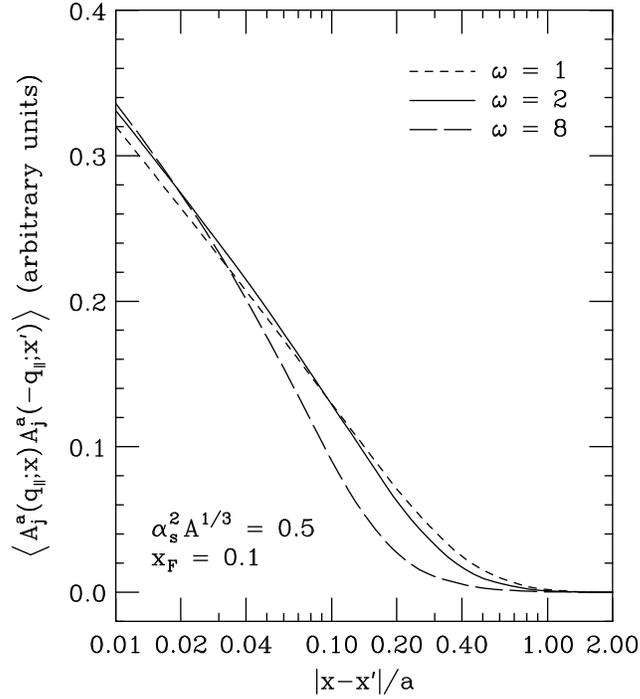}

\caption[]{
Position space correlation functions used to determine
the fully differential gluon number density from 
Eq.~(\protect\ref{x1629}) evaluated using the
power-law model for $\widetilde{\DIII}(\qIII\ts)$ given in 
Eq.~(\ref{PowerLaw}).  The three curves compare the results
using $\omega = 1$, 2, and 8 at fixed $\xf = 0.1$ in the
Abelian limit ($\alpha_s^2 A^{1/3} = 0$) and for uranium
($\alpha_s^2 A^{1/3} = 0.5$).
}
\label{OmegaDep}
\end{figure}

In this section we will illustrate the features of the gluon
number density described in the previous section
by choosing a specific form for $\widetilde{\DIII}(\qIII\ts)$,
namely
\beq
\widetilde{\DIII}(\qz;\qt) \equiv
1 -
{ {1}\over{[1+\aw^2(\qt^2+\qz^2)]^{\omega}} },
\label{PowerLaw}
\eeq
where $\omega$ is an arbitrary positive integer,
and $\aw \equiv a/\sqrt{3\omega}$.
This function obviously satisfies the color neutrality
condition~(\ref{CNmom}).
In terms of the Kovchegov model\cite{paper12},
the choice made in Eq.~(\ref{PowerLaw}) corresponds
to a Yukawa-like distribution of quarks within the nucleon
(see Sec.~\ref{KM}).  The value of $\aw$ has been
chosen so that the root-mean-square radius of the nucleon
in precisely $a$.
Eq.~(\ref{PowerLaw}) is also convenient in that we are
able to perform the integrals in Eq.~(\ref{x1490}) and~(\ref{x1488a})
analytically (see Appendix~\ref{PowerDetails}):
\beqa
\curlyL(\xf;\deltat) =
&-& {{\omega}\over{4\pi}}\ts (\xf\mn\aw)^2 K_0(\xf\mn\Delta)
\phantom{\Biggl[}
\cr &+&
{{1}\over{4\pi}}
\sum_{j=0}^{\omega-1}
{{1}\over{j!}} \ts
\biggl( {{\Delta}\over{2\aw}} \biggr)^j \ts
{
{  1 + (\omega{-}j)(\xf\mn\aw)^2  }
\over
{ \Bigl[ 1 + (\xf\mn\aw)^2 \Bigr]^{j/2} }
} \ts
K_{j} \biggl( {{\Delta}\over{\aw}}
\sqrt{ 1 + (\xf\mn\aw)^2 } \ts\biggr),
\label{multipole-curlyL}
\eeqa
and
\beqa
L(\deltat) =
&-& {{\aw^2}\over{2\pi}}\ts\omega
\Biggl[ K_0 \biggl( {{\Delta}\over{\aw}} \biggr)
+ \ln\biggl( {{\Delta}\over{2\aw}} \biggr)
+ \gamma_E \ts
\Biggr]
\cr &+&
{{\aw^2}\over{2\pi}} 
\sum_{j=1}^{\omega-1}
(\omega{-}j) \ts
\Biggl[
{{1}\over{2j}} -
{{1}\over{j!}} \ts
\biggl( {{\Delta}\over{2\aw}} \biggr)^j \ts
K_{j} \biggl( {{\Delta}\over{\aw}} \biggr)
\Biggr].
\label{multipole-plainL}
\eeqa
In Fig.~\ref{OmegaDep} we have plotted the integrand of
the gluon number density~(\ref{x1629}) (omitting the exponential
factor) for various values of $\omega$:  according to
Eq.~(\ref{gnum3d}) this is just proportional to
$\langle A_i^a(\qz;\xt) A_i^a(-\qz;\xprimet) \rangle$.
We see from the plots that the
non-Abelian corrections become more important as
$\omega$ increases:  the range of the (position space)
integrand decreases.  Thus, we would expect to find fewer soft
gluons in a model with larger $\omega$.  Overall, however, the
dependence on $\omega$ is rather weak.
Therefore, we have chosen to present plots only for the $\omega=1$
case for the rest of this discussion.

\begin{figure}[h]

\vspace*{20cm}
\includegraphics{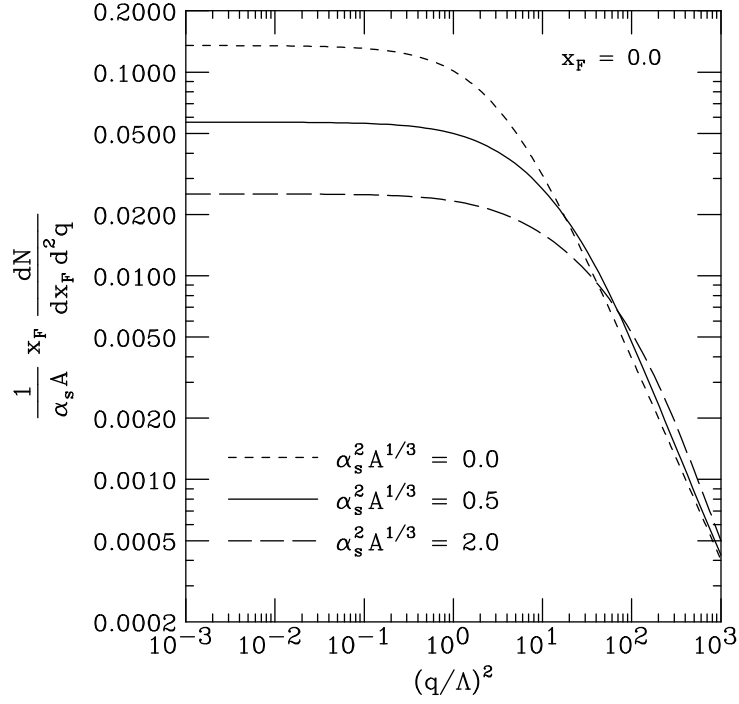}
\includegraphics{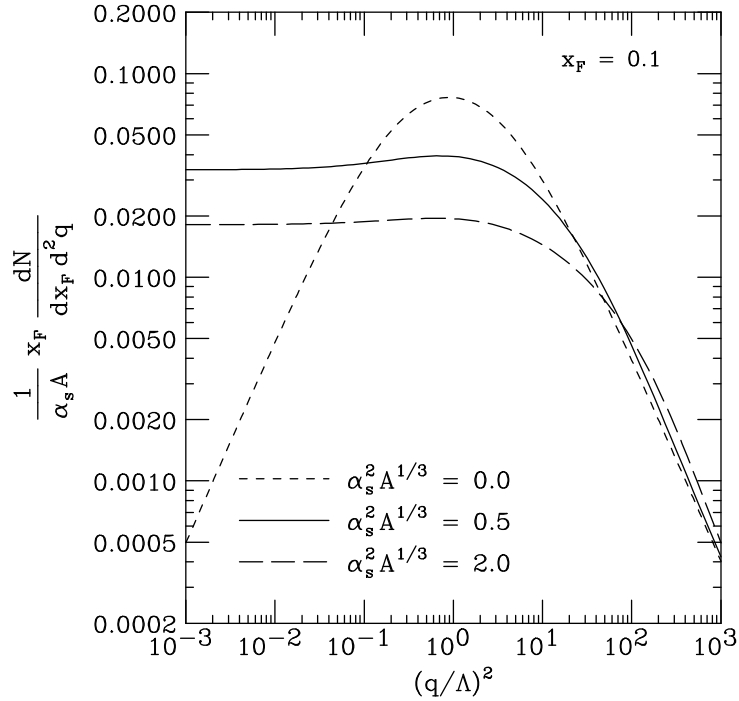}

\caption[]{The fully differential gluon number density
Eq.~(\protect\ref{x1629}) evaluated in the power-law model
with $\omega=1$.  The three curves on each plot represent
the Abelian limit ($\alpha_s^2 A^{1/3} = 0$), uranium
($\alpha_s^2 A^{1/3} = 0.5$), and a large toy nucleus
with $A\sim 15000$ ($\alpha_s^2 A^{1/3} = 2.0$).
The upper plot illustrates the results for $\xf\rightarrow 0$,
whereas the lower plot has been drawn for $\xf = 0.1$.
}
\label{x1282C}
\end{figure}

\begin{figure}[h]

\vspace*{19.2cm}
\includegraphics{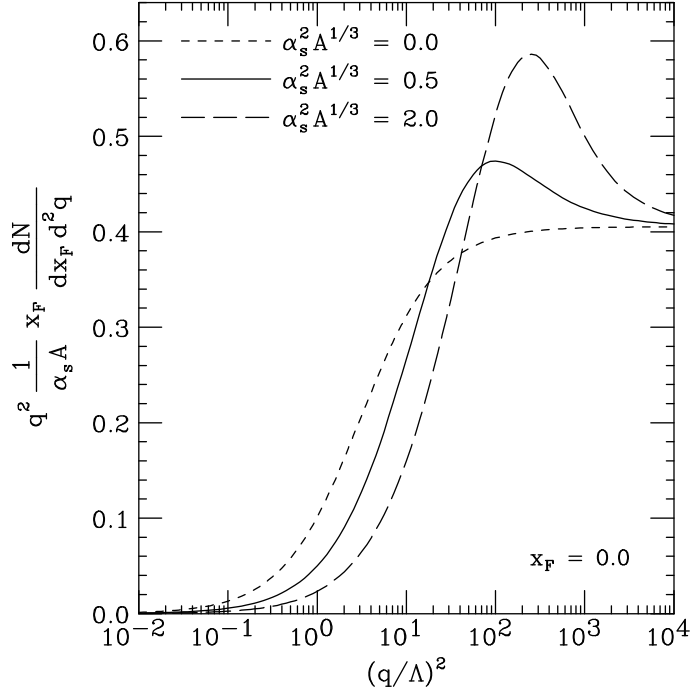}
\includegraphics{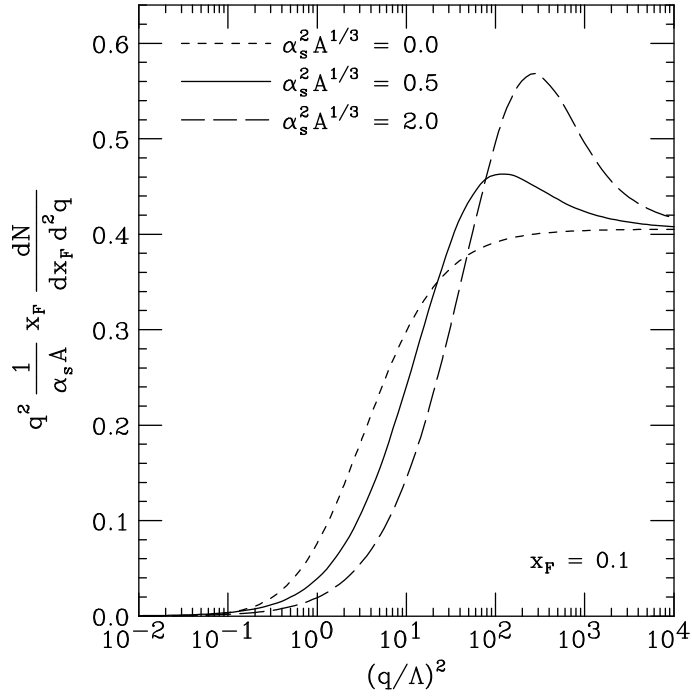}

\caption[]{The fully differential gluon number density
Eq.~(\protect\ref{x1629}) multiplied by $\qt^2$ and
evaluated in the $\omega=1$ power-law model.
These plots accurately reflect the relative contributions
to the gluon structure function $g_A(\xf,Q^2)$ coming
from each value of $\qt^2$.
The three curves on each plot represent
the Abelian limit ($\alpha_s^2 A^{1/3} = 0$), uranium
($\alpha_s^2 A^{1/3} = 0.5$), and a large toy nucleus
with $A\sim 15000$ ($\alpha_s^2 A^{1/3} = 2.0$).
The upper plot illustrates the results for $\xf\rightarrow 0$,
whereas the lower plot has been drawn for $\xf = 0.1$.
}
\label{ksqrd-plot}
\end{figure}

Fig.~\ref{x1282C} contains plots of the fully differential gluon number
density as a function of $\qt^2$ at 
$\xf = 0.0$ and 0.1.\footnote{By $\xf = 0.0$ we really
mean some value of $\xf$ in the MV region, $\xf \ll A^{-1/3}/(\mn a)$.} 
Three different values of $\alpha_s^2 A^{1/3}$ have been used,
namely 0.0, 0.5, and 2.0, corresponding to the Abelian limit,
a (roughly) uranium-sized nucleus, and a very large (toy) nucleus. 
In all cases the distributions saturate as $\qt^2 \rightarrow 0$,
with the turn-over occurring at a few times $\LQCD^2$.
This turn-over is very much like the one 
which Mueller sees in his calculation based upon
onium-scattering\cite{paper19,paper37}.
In each plot, the maximum value reached
by $\xf dN/d\xf d^2\qt$ decreases as $A^{1/3}$ is increased.
At large $\qt^2$, the distributions
exhibit the $1/\qt^2$ fall-off characteristic of individual
point charges.
Fig.~\ref{ksqrd-plot} illustrates the same fully differential
gluon number densities, but multiplied by a factor of $\qt^2$.
These plots are useful because the visual area under the curves
(using  logarithmic horizontal  and linear vertical scales)
faithfully reproduces the result of the integration defining
the gluon structure function~(\ref{StrFnDef}):  what-you-see
is what-you-get.  From these plots, we see that the very small
$\qt^2$ region makes very little contribution to $g_A(\xf,Q^2)$.
At very large $\qt^2$ all of the curves converge to the same
result, as required by the transverse momentum sum rule.
We also see a pile-up of gluons in the region of a few times $\LQCD^2$.
As the size of the nucleus is increased, this peak shifts to
larger $\qt^2$ and increases in size.  In Fig.~\ref{PeakPlot} we 
track the location of this peak as a function of $v^2$.
This plot clearly shows that for large-enough nuclei, 
our heuristic argument of Sec.~\ref{GENERAL} claiming that
$Q^2_{peak}$ ought to be proportional to $v^2$ 
is very nearly correct, with a 
proportionality constant close to unity.
When $A^{1/3}$ is too small, however, this relation breaks down
as the scale associated with $\curlyE(v^2L(\deltat))$ no
longer dominates the result.

\begin{figure}[h]

\vspace*{17cm}
\includegraphics{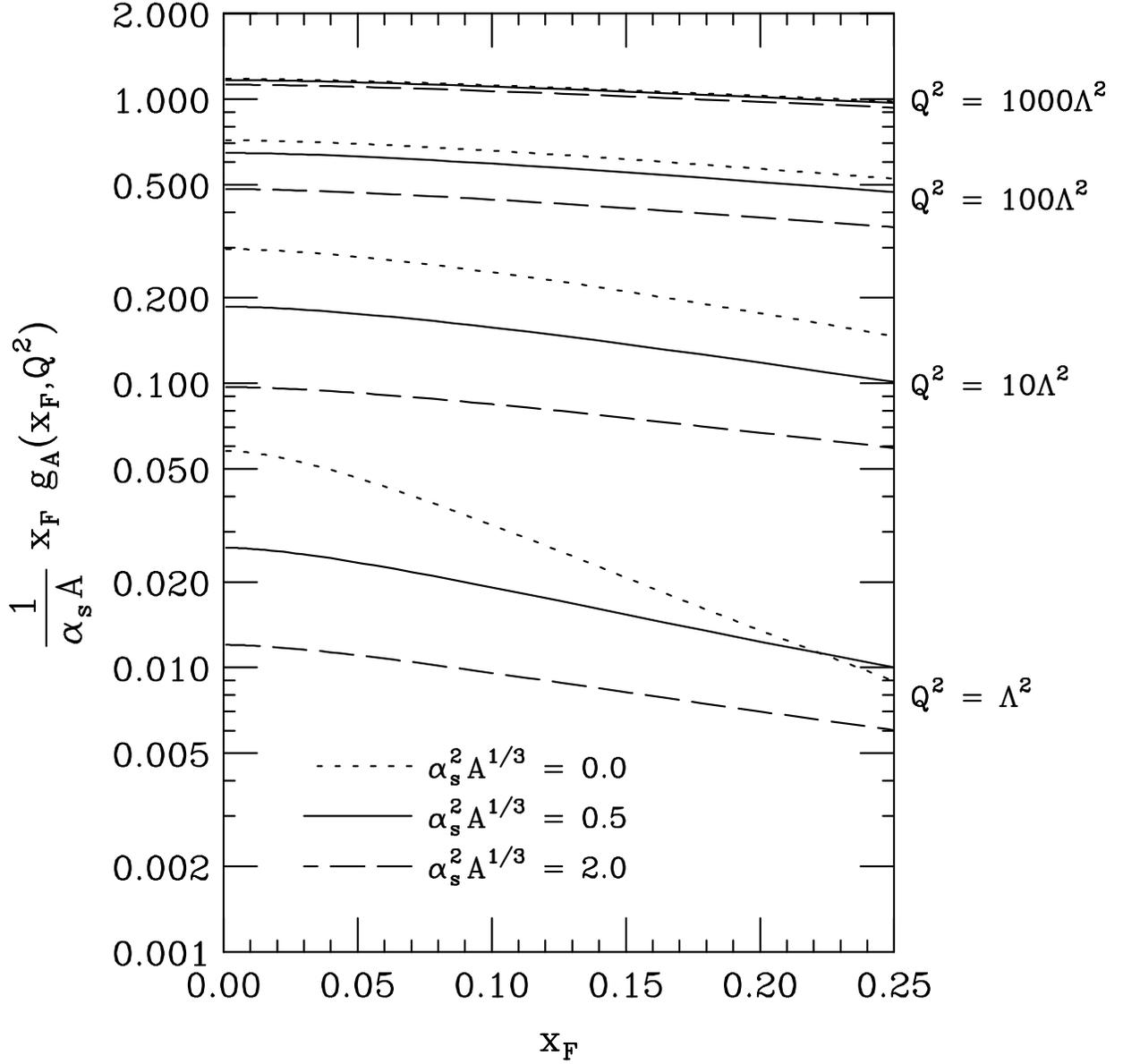}

\caption[]{Gluon distribution function $g_A(\xf,Q^2)$
in the $\omega=1$ power-law model
plotted versus $\xf$ for $Q^2 = \LQCD^2$, $10\LQCD^2$, $100\LQCD^2$,
and $1000\LQCD^2$. 
The three curves at each $Q^2$ value are for
the Abelian limit ($\alpha_s^2 A^{1/3} = 0$), uranium
($\alpha_s^2 A^{1/3} = 0.5$), and a large toy nucleus
with $A\sim 15000$ ($\alpha_s^2 A^{1/3} = 2.0$).
}
\label{x1286B}
\end{figure}

In Fig.~\ref{x1286B} we plot the gluon structure function
per nucleon as a function of $\xf$  for various values of $Q^2$.
Although we have drawn all of these curves over the entire range
from $\xf=0$ to $\xf=0.25$,
we remind the reader that the maximum $\xf$
value for which our calculation can be trusted decreases as $Q^2$
is increased (see Fig.~\ref{ValidRegion}).
Away from the very small $\xf$ region, we see that as
$\xf$ increases, $\xf g_A(\xf,Q^2)$ decreases, indicating a
fall off which is more rapid than $1/\xf$.  The degree of
dependence on $A^{1/3}$ goes down as $Q^2$ is increased:
at $Q^2 = 1000\LQCD^2$ there is very little nuclear dependence
beyond the trivial scaling with the number of nucleons.

\begin{figure}[h]

\vspace*{17cm}
\includegraphics{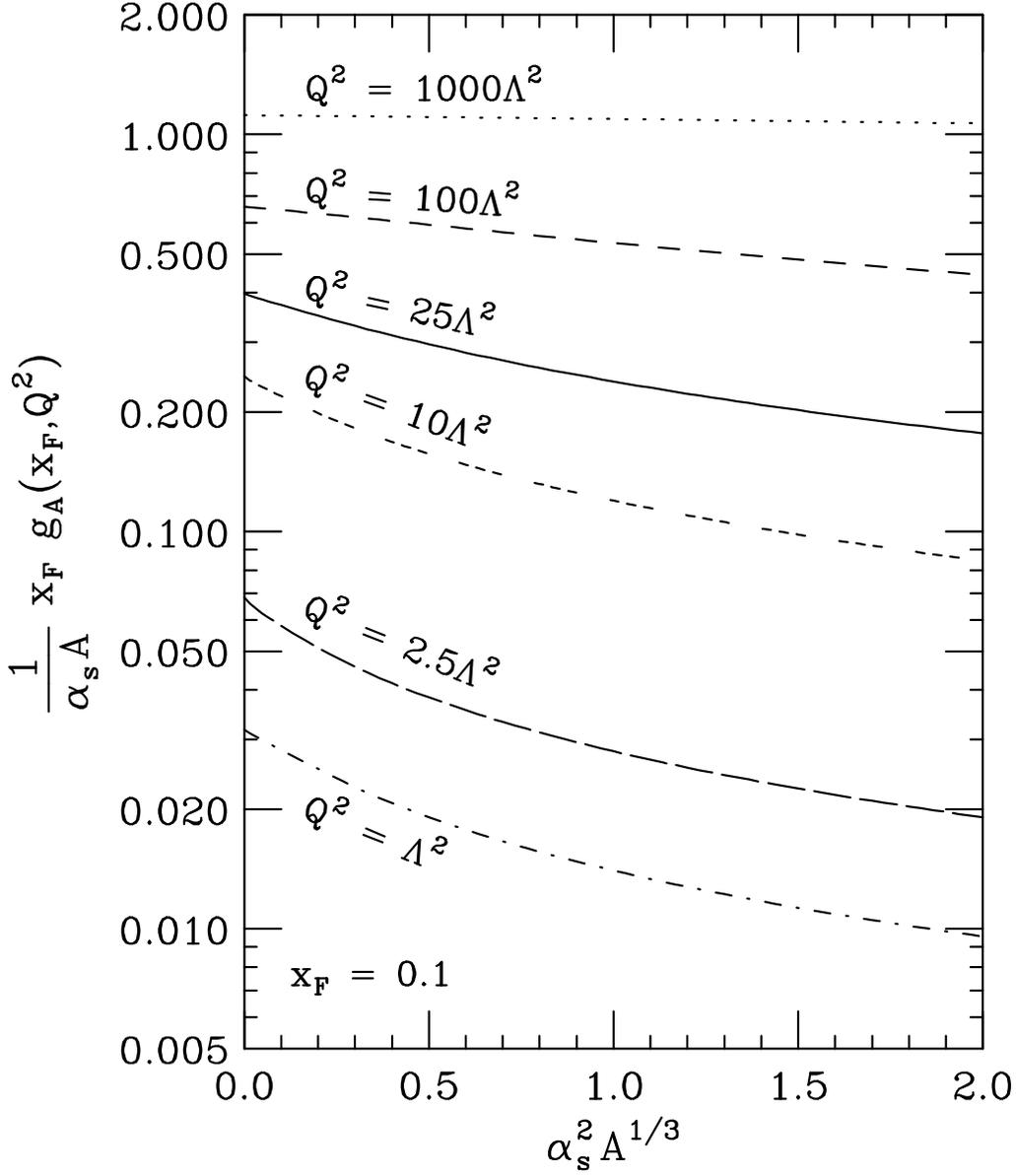}

\caption[]{Nuclear dependence of the gluon distribution function 
$g_A(\xf,Q^2)$ in the $\omega=1$ power-law model for fixed $\xf = 0.1$
and
$Q^2 = \LQCD^2$, $2.5\LQCD^2$, $10\LQCD^2$, $25\LQCD^2$, $100\LQCD^2$,
and $1000\LQCD^2$.  These functions grow more slowly than $A$
as the number of nucleons is increased.
}
\label{CHIPLOT}
\end{figure}

In Fig.~\ref{CHIPLOT} we further explore the nuclear dependence of our
result, by plotting the gluon structure function per nucleon
as a function of $\alpha_s^2 A^{1/3}$ at fixed $\xf = 0.1$
and several different values of $Q^2$.
At low $Q^2$, we see a marked departure from the na{\"\i}ve
expectation:  the number of gluons is not simply proportional
to the number of nucleons.   Instead, our gluon
structure function grows more slowly than $A$ as the number
of nucleons is increased.
At larger values of $Q^2$, however,
the nuclear dependence is reduced as the non-Abelian
corrections become less important.


\subsection{Connection to the Kovchegov Model}\label{KM}

In this subsection we put our power-law model into the context
of the Kovchegov model of Ref.~\cite{paper12}.
In this model we imagine a nucleus of radius $R$ which contains
$A$ nucleons, each of radius $a \sim \LQCD^{-1}$.
Each of the ``nucleons'' is made up of a quark-antiquark pair.
Under these assumptions, Ref.~\cite{paper12} provides a means
for computing the two-point charge density correlation function.
The correlation function derived in Ref.~\cite{paper12}
is applicable to the 2-dimensional case.  However, since the
2-dimensional form was arrived at from a 3-dimensional one
by integrating over the longitudinal variables, it is not
difficult to modify the derivation of Ref.~\cite{paper12} 
to cover the 3-dimensional case we are studying here.
Parameterizing the correlation function as in Eq.~(\ref{rhorho3})
we find that
\beq
{{1}\over{V}} \ts
{\SIII}\biggl( {{\xIII+\xIII\ts'}\over{2}} \biggr)
{\DIII}(\xIII-\xIII\ts') \equiv
\Isng (\xIII,\xprimeIII) - \Ismth(\xIII,\xprimeIII),
\label{Kconnection}
\eeq
where
\beq
\Isng(\xIII,\xprimeIII) =
\int d^3 \rIII\ts\ts \vert \phi(\rIII\ts)\vert^2
\int d^3 \xiIII\ts\ts \vert \psi(\xiIII\ts)\vert^2
\ts \delta^3(\xIII-\rIII-\xiIII\ts) \ts
\delta^3(\xprimeIII-\rIII-\xiIII\ts)
\label{Isng}
\eeq
and
\beq
\Isng(\xIII,\xprimeIII) =
\int d^3 \rIII\ts\ts \vert \phi(\rIII\ts)\vert^2
\int d^3 \xiIII\ts\ts \vert \psi(\xiIII\ts)\vert^2
\int d^3 \xiprimeIII\ts\ts \vert \psi(\xiprimeIII)\vert^2
\ts \delta^3(\xIII-\rIII-\xiIII\ts) \ts
\delta^3(\xprimeIII-\rIII-\xiprimeIII).
\label{Ismth}
\eeq
In Eqs.~(\ref{Isng}) and~(\ref{Ismth}) the position of the
nucleon relative to the center of the nucleus is denoted by $\rIII$
while the position of the (anti)quark relative to the center
of the nucleon is denoted by $\xiIII\ts^{(\prime )}$.
In addition to making the result fully 3-dimensional, we have
allowed for the possibility that the quarks and nucleons have
some distribution other than uniform:  the function
$\vert\phi(\rIII\ts)\vert^2$ gives the probability distribution
for nucleons within the nucleus while 
$\vert\psi(\xiIII\ts)\vert^2$ does the same for the (anti)quarks 
within the nucleons.
Both functions are normalized to unit total integral.
If we compute them exactly from Eqs.~(\ref{Isng}) and~(\ref{Ismth}),
the functions $\Isng(\xIII,\xprimeIII)$ and $\Ismth(\xIII,\xprimeIII)$
do not factorize on the sum and difference variables as implied
by the left hand side of Eq.~(\ref{Kconnection}).  However,
if we approximate the integrals by assuming that $a\ll R$
(the large nucleus approximation), then they do factorize as
advertised, and we obtain
\beq
\SIII(\sigmaIII\ts) =
V \vert \phi(\sigmaIII\ts) \vert^2;
\label{KovS}
\eeq
\beq
\DIII(\deltaIII\ts) = \delta^3(\deltaIII\ts)
- \int d^3 \xiIII\ts\ts \vert \psi(\xiIII\ts)\vert^2
  \int d^3 \xiprimeIII\ts \vert \psi(\xiprimeIII)\vert^2
\delta^3(\xiIII-\xiprimeIII-\deltaIII\ts).
\label{KovD}
\eeq
The separation of the 
right hand side of~(\ref{Kconnection}) into  $\SIII$ and $\DIII$
has been uniquely fixed by imposing the normalization 
for $\SIII$ specified in Eq.~(\ref{Snorm}).

\begin{figure}[h]

\vspace*{17cm}
\includegraphics{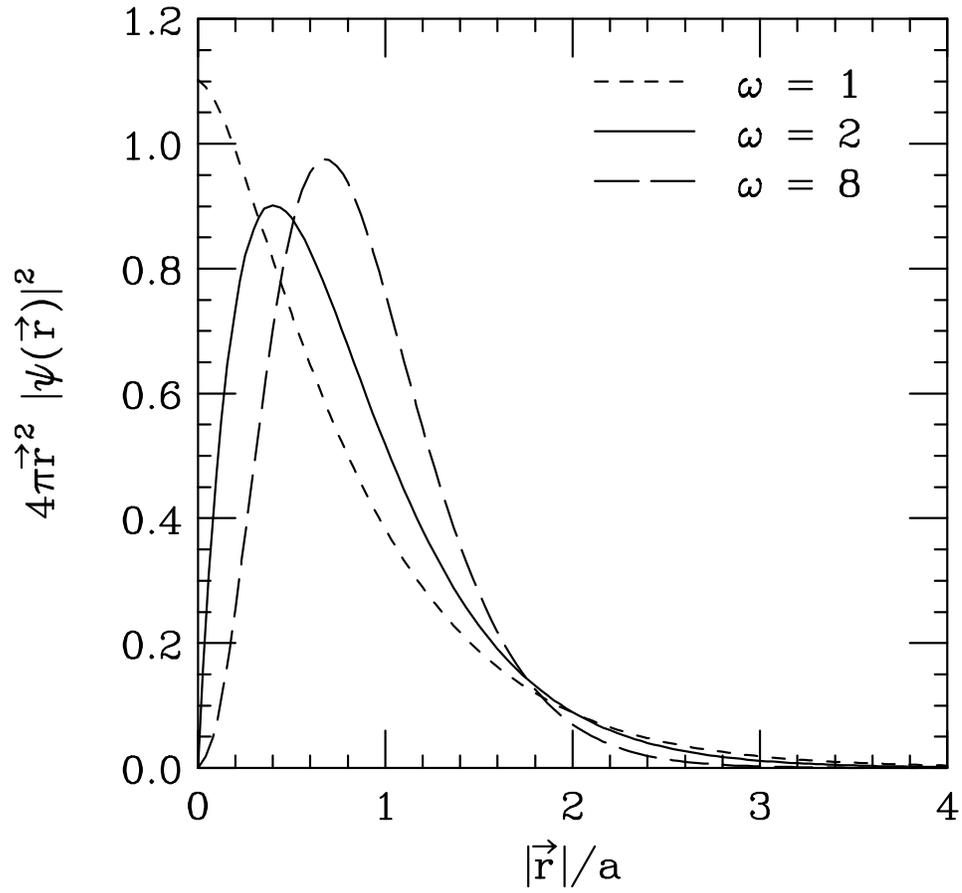}

\caption[]{Quark probability distributions as determined in the
context of the Kovchegov model\protect\cite{paper12} 
for the power-law correlation function~(\protect\ref{PowerLaw}).
The three curves compare the results for $\omega=1$, 2, and 8.
}
\label{PSI2}
\end{figure}

To make the connection with our power-law model, we first
Fourier transform Eq.~(\ref{KovD}) to obtain
\beq
1 - \widetilde\DIII(\qIII\ts) =
\Biggl\vert
\int d^3 \xIII\ts\ts e^{i\qIII\cdot\xIII} \vert\psi(\xIII\ts)\vert^2
\Biggr\vert^2.
\label{x1538}
\eeq
For a spherically-symmetric nucleus, $\widetilde\DIII(\qIII\ts)$
is real.  Thus, Eq.~(\ref{x1538}) is telling us
that the square root of $1{-}\widetilde\DIII$ is given by the
Fourier transform of $\vert\psi\vert^2$.  Consequently
\beq
\vert\psi(\xIII\ts)\vert^2 =
\int {{d^3\qIII}\over{8\pi^3}} \ts
e^{-i\qIII\cdot\xIII} 
\sqrt{1-\widetilde\DIII(\qIII\ts)}.
\label{x1545}
\eeq
For the power-law model~(\ref{PowerLaw}), the integral 
in Eq.~(\ref{x1545})
is easily performed using the methods 
outlined in Appendix~\ref{PowerDetails},
with the result
\beq
\vert \psi(\xIII\ts) \vert^2 =
{ {1}\over{4(\pi \aw^2)^{3/2}} }
\ts
{ {1}\over{\Gamma(\omega/2)} }
\ts
\biggl(
{ {\xt^2 + \xz^2}\over{4\aw^2} }
\biggr)^{{\omega-3}\over{4}}
K_{{\omega-3}\over{2}}
\biggl(
{{1}\over{\aw}}
\sqrt{\xt^2 + \xz^2} \ts
\biggr).
\label{x1547}
\eeq
Eq.~(\ref{x1547}) describes a Yukawa-like distribution of
quarks within the nucleons in the sense that the long-distance
behaviour of the modified Bessel function is a dying
exponential:
\beq
4\pi\xIII\ts^2
\vert \psi(\xIII\ts) \vert^2 \ts
\mathop{\longrightarrow}_{\scriptstyle{\vert\xIII\vert\rightarrow\infty}} 
\ts { {2}\over{\aw} }
\ts
{ {1}\over{\Gamma(\omega/2)} }
\ts
\biggl(
{ {\xt^2 + \xz^2}\over{4\aw^2} }
\biggr)^{{\omega}/{4}}
\exp\biggl( -{{1}\over{\aw}}
\sqrt{\xt^2 + \xz^2}\ts 
\biggr).
\label{x1547B}
\eeq
At the origin, we have
\beq
4\pi\xIII\ts^2
\vert \psi(\xIII\ts) \vert^2 \ts
\mathop{\longrightarrow}_{\scriptstyle{\vert\xIII\vert\rightarrow0}} 
\ts
\cases{ 
constant,
                       & $\omega=1$; \cr
\phantom{x}& \cr
0,
                       & $\omega \ge 2$. \cr
}
\eeq
These distributions are plotted 
in Fig.~\ref{PSI2}
for $\omega=1$, 2, and 8.


\section{Conclusions}\label{CONCLUSIONS}

We have extended the McLerran-Venugopalan model for the
gluon distribution of a very large nucleus to larger values
of $\xf$.
The classical computation contained in this paper, like those in
Refs.~\cite{paper1,paper2,paper3,paper4,paper9,PAPER14}, is
based on the premise that the
quantum corrections may be ignored since
we are working in a regime where $\alpha_s \ll 1$.
However, at small $\xf$ there are 
quantum corrections to the distribution functions of the 
MV model which are proportional to
$\alpha_s\ln(1/\xf)$\cite{paper4,paper7}.
For fixed $\qt^2 \sim \LQCD^2$, it can be shown that 
$\alpha_s \ln(1/\xf)$ is of order 1 in region C of 
Fig.~\ref{ValidRegion}, and that furthermore, $\alpha_s\ln(1/\xf)$
increases as $\xf$ is decreased into regions A and B.
This suggests that the quantum corrections may be more manageable
in region C:  a leading-log calculation to resum these contributions
might suffice.  Considerable 
effort\cite{paper9,paper34,paper11,paper39,paper40,paper43,paper44}
has already been performed with the goal of
dealing with the quantum corrections in the very small
$\xf$ region where $\alpha_s\ln(1/\xf) \gg 1$.

The values of $\xf$ we have considered in this paper are in
the regime where the gluons begin to probe the
longitudinal structure of the nucleus.  A description of the
physics in this situation must begin with a fully three-dimensional
framework and a source which is slightly off of the light cone.
We have solved the Yang-Mills equations for such a source in
the covariant gauge to lowest order in $g^2$, and then transformed
that solution for the vector potential to the light-cone gauge,
where the connection to the gluon number density is to be made.
The determination of the gluon number density itself relies
heavily on the
fact that we are considering a large nucleus ($R \gg a$)
which consists of color-neutral nucleons.
We have obtained a relatively compact expression in this limit,
which sums the non-Abelian effects to all orders in 
$\alpha_s^2 A^{1/3}$.
For $\xf\rightarrow 0$, our results match smoothly onto the
previous treatments\cite{paper9,PAPER14}.  
Our results for the gluon number density exhibit saturation
at small $\qt^2$:  instead of diverging as $\qt^2 \rightarrow 0$,
the distributions approach a finite constant,
as illustrated in Fig.~\ref{x1282C}.
The nuclear corrections induce a pile-up of gluons at
$\qt^2 \sim \vsqr$, where $\vsqr \propto A^{1/3}\LQCD^2$.
In addition, the gluon structure functions which we obtain
grow less rapidly than $A$ as the number of nucleons is 
increased, as illustrated in Fig.~\ref{CHIPLOT}.


\acknowledgements

This research is supported in part by
the Natural Sciences and Engineering Research Council of Canada
and the Fonds pour la Formation de Chercheurs 
et l'Aide \`a la Recherche of Qu\'ebec.
We would like to thank Sangyong Jeon for several insightful
discussions during the course of this work.



\appendix


\section{Explicit Computation of Gluon Number Density}\label{CONTRACTIONS}

In this Appendix we will present the details of the derivation 
leading to the expression for the gluon number density
presented in Eqs.~(\ref{x1629})--(\ref{x1629V}).
Our starting point is the light-cone gauge expression
for the vector potential:
\beqa
A^{j}(\qz;\xt) = 
{{1}\over{ig\qz}} 
\sum_{m=1}^{\infty} 
(ig^2)^{m}   &&
\int_{-\infty}^{\infty} d^m{\yz}_{\downarrow} 
\int_{-\infty}^{\infty} d^m\xiz
\ts \exp(-i\qz{\yz}_1)
\cr \times &&
\int {{d^2\pt_1}\over{4\pi^2}} \ts
{{p_{1j}}\over{2p_1}} \ts
e^{-i\pt_1\cdot\xt} 
\exp\Bigl(-p_1\vert{\yz}_1-{\xiz}_1\vert\Bigr)
\cr  \times &&
\prod_{\ell=2}^{m}
\int {{d^2\pt_\ell}\over{4\pi^2}} \ts
{{1}\over{2p_\ell}} \ts
e^{-i\pt_\ell\cdot\xt} 
\exp\Bigl(-p_\ell\vert{\yz}_\ell-{\xiz}_\ell\vert\Bigr)
\cr \times &&
\Lbrk
\rho({\xiz}_1;\pt_1)
\rho({\xiz}_2;\pt_2) \cdots 
\rho({\xiz}_m;\pt_m)
\Rbrk. \phantom{\biggl[}
\label{x1609-APPX}
\eeqa
Simply inserting two copies of Eq.~(\ref{x1609-APPX}) into
the master formula~(\ref{gnum3d})
leaves us with a bewildering array of contractions which must be
performed.  
A more efficient way to proceed is to organize the computation
in two stages, as was done in Ref.~\cite{paper9}.
First, we consider all of the
ways in which pairs of $\rho$'s within a single $A^j(\qz;\xt)$ 
may be contracted.  
We will see that these self-contractions exponentiate
provided that $a/R \ll 1$.
Afterwards, we will deal with the mutual contractions, where one
$\rho$ comes from each factor of $A^j(\qz;\xt)$.
In the end, we retain exactly the same terms as were retained
in Ref.~\cite{paper9}.  However, the justification for keeping
only these terms is very different from the one in Ref.~\cite{paper9},
and relies heavily on the large nucleus approximation, $a/R \ll 1$.

Since Eq.~(\ref{x1609-APPX}) contains charge densities which have been
Fourier-transformed on the transverse variables, it is convenient
to do the same to the two point correlation function.  It is 
straightforward to show that the 
corresponding transform of Eq.~(\ref{rhorho3})
reads
\beq
\Bigl\langle 
\rho^a(\xiz;\pt) \rho^b(\xiz';\pt') 
\Bigr\rangle \equiv
\delta^{ab} \Kcube \ts
{\SIII}\Bigl( \half(\xiz+\xiz'); \pt+\pt' \Bigr)
{\DIII}\Bigl(\xiz-\xiz';\half(\pt-\pt') \Bigr) .
\label{mixed-rhorho}
\eeq


\subsection{Self Contractions}

In this subsection, we will show that the various non-vanishing
self-contractions within a single $A^j(\qz;\xt)$ may be arranged 
into an exponential factor.
We will use the normal-ordered notation $: \cdots :$ as a
bookkeeping device to indicate which factors are not to undergo 
further self-contractions.

We begin with the observation that the contraction
between $\rho({\xiz}_1;\pt_1)$ and
$\rho({\xiz}_2;\pt_2)$ vanishes identically:
Eq.~(\ref{mixed-rhorho})
is symmetric in the color indices, 
whereas the factors being contracted 
are antisymmetric, appearing in the innermost commutator.

Next, we show that
in the limit $a/R \ll 1$ ({\it i.e.} $A^{1/3} \gg 1$), 
the only non-vanishing self-contractions
are between adjacent factors of $\rho$
(such as the self-contraction illustrated in Fig.~\ref{EighthOrder}b).
Suppose we consider a term in which we contract the
non-adjacent factors
$\rho({\xiz}_i;\pt_i)$ with $\rho({\xiz}_j;\pt_j)$, where
$j>i+1$:  an example of such a contraction is illustrated in
Fig.~\ref{EighthOrder}a.
The relevant longitudinal factors coming
from this type of contribution read\footnote{The additional
factor of $\exp(-i\qz{\yz}_1)$
which would also be present if $i=1$ does not affect the outcome
of our argument.}
\beqa
\int_{-\infty}^{\infty}  d{\yz}_i 
\int_{-\infty}^{\infty} d{\yz}_j \ts  &&
\Theta({\yz}_{i-1}-{\yz}_i) \ts
\Theta({\yz}_{i}-{\yz}_{i+1}) \ts
\Theta({\yz}_{j-1}-{\yz}_j) \ts
\Theta({\yz}_{j}-{\yz}_{j+1}) \ts
\cr \times  &&
\int_{-\infty}^{\infty} d{\xiz}_i
\int_{-\infty}^{\infty} d{\xiz}_j  \ts
\exp\Bigl(-p_i \vert{\yz}_i-{\xiz}_i \vert \Bigr)
 \exp\Bigl(-p_j \vert{\yz}_j-{\xiz}_j \vert \Bigr)
\cr && \qquad\qquad\qquad\qquad\nts \times
 {\SIII}\Bigl( \half({\xiz}_i+{\xiz}_j);\pt_i+\pt_j \Bigr)
 {\DIII}\Bigl({\xiz}_i-{\xiz}_j;\half(\pt_i-\pt_j)\Bigr),
\label{x1526}
\eeqa
where the $\Theta$-functions encode the ordering required in
the $\yz$ integrations.
To see how this contribution is subleading when we make the large
nucleus approximation, we first introduce
\beq
{\uz}_i \equiv {\yz}_i - {\xiz}_i; \qquad
{\uz}_j \equiv {\yz}_j - {\xiz}_j
\label{vchange1}
\eeq
in favor of ${\xiz}_i$ and ${\xiz}_j$, followed by 
\beq
\sigmaz \equiv \half({\yz}_i {-} {\uz}_i {+} {\yz}_j {-} {\uz}_j); 
\qquad
\deltaz \equiv {\yz}_i {-} {\uz}_i {-} {\yz}_j {+} {\uz}_j
\label{vchange2}
\eeq
to replace ${\yz}_i$ and ${\yz}_j$.  The Jacobians of both 
transformations are unity.  Hence,~(\ref{x1526}) becomes
\beqa
\int_{-\infty}^{\infty} \nts d\sigmaz
\int_{-\infty}^{\infty} \nts d\deltaz
\int_{-\infty}^{\infty} \nts d{\uz}_i
\int_{-\infty}^{\infty} \nts d{\uz}_j \ts &&
\Theta({\yz}_{i-1} {-} \sigmaz {-} \half\deltaz {-} {\uz}_i) \ts
\Theta(\sigmaz {+} \half\deltaz {+} {\uz}_i {-} {\yz}_{i+1}) \ts
\cr \times &&
\Theta({\yz}_{j-1} {-} \sigmaz {+} \half\deltaz {-} {\uz}_j) \ts
\Theta(\sigmaz {-} \half\deltaz {+} {\uz}_j {-} {\yz}_{j+1}) \ts
\cr \phantom{\biggl(} \times && 
\exp \Bigl( -p_i\vert {\uz}_i \vert {-} p_j \vert {\uz}_j \vert \Bigr)
\cr \phantom{\biggl(} \times && 
 {\SIII}(\sigmaz;\pt_i+\pt_j)
 {\DIII}\Bigl(\deltaz;\half(\pt_i-\pt_j)\Bigr),
\label{x1527}
\eeqa
Recall that the function $\SIII(\sigmaIII\ts)$ is nonvanishing 
provided that $\vert\sigmaIII\ts\vert \alt R$.
Likewise, the function $\DIII(\deltaIII\ts)$ is dominated by the
region where $\vert\deltaIII\ts\vert \alt a$.
Therefore, as far as the transverse integrations are concerned,
the integrand in~(\ref{x1527}) 
is dominated by the region $\sigmaz \gg \deltaz$.
Neglect of $\deltaz$ relative to $\sigmaz$ in the
$\Theta$-functions will result in errors of order $a/R$.
Furthermore, the exponential factor restricts the values of
${\uz}_i$ and ${\uz}_j$ for which the integrand is significant
to ${\uz}_i, {\uz}_j \alt p_i, p_j$.  So unless the region where
$p_i$ or $p_j$ is $\alt 1/a$ is important,
we may also drop ${\uz}_i$ and ${\uz}_j$
from the $\Theta$-function arguments.  
However, we know that the typical momenta associated with
$\SIII(\sigmaz;\pt_i+\pt_j)$
are $\vert\pt_i+\pt_j\vert \sim 1/R$, whereas the
typical momenta associated with
${\DIII}\Bigl(\deltaz;\half(\pt_i-\pt_j)\Bigr)$
are $\half\vert\pt_i-\pt_j\vert \sim 1/a$.
Together, these constraints imply that the main contributions
to the integral occur when
$\pt_i$ and $\pt_j$ are back-to-back
to within an amount of order $1/R$, and they each posses a 
magnitude of order $1/a$.
Thus we conclude that
${\uz}_i$ and ${\uz}_j$ are indeed of order $a$, and
may be dropped from the $\Theta$-functions.
Making these approximations in~(\ref{x1527}) yields
\beqa
\int_{-\infty}^{\infty} \nts\nts d\sigmaz
\int_{-\infty}^{\infty} \nts\nts d\deltaz
\int_{-\infty}^{\infty} \nts\nts d{\uz}_i &&
\int_{-\infty}^{\infty} \nts\nts d{\uz}_j  \ts 
\Theta({\yz}_{i-1} {-} \sigmaz) \ts
\Theta(\sigmaz {-} {\yz}_{i+1}) \ts
\Theta({\yz}_{j-1} {-} \sigmaz) \ts
\Theta(\sigmaz{-}{\yz}_{j+1}) 
\cr \phantom{\biggl(}  &&  \times
\exp \Bigl( -p_i\vert {\uz}_i \vert {-} p_j \vert {\uz}_j \vert \Bigr)
 {\SIII}(\sigmaz;\pt_i+\pt_j)
 {\DIII}\Bigl(\deltaz;\half(\pt_i-\pt_j)\Bigr).
\label{x1527a}
\eeqa
The $\Theta$-functions in this expression tell us that $\sigmaz$
should lie between ${\yz}_{i-1}$ and ${\yz}_{i+1}$ on one hand,
and (simultaneously) lie between ${\yz}_{j-1}$ and ${\yz}_{j+1}$
on the other.
However, the $\yz$'s are ordered, and since we are considering
non-adjacent factors, $j > i+1$, these to ranges do not overlap.
Thus,~(\ref{x1527a}) vanishes, and we conclude that the contributions
from contraction between non-adjacent $\rho$'s are suppressed
by one or more powers of $a/R$ relative to contributions
from contractions between adjacent $\rho$'s.

Now that we know which self-contractions may be ignored, let us begin
a term-by-term examination of the series in Eq.~(\ref{x1609-APPX}).
We will denote the $m$th term in the sum by 
$\Ait{j}{m}(\qz;\xt)$.
The first term, $\Ait{j}{1}(\qz;\xt)$, has only a single factor
of $\rho$.  Thus, we trivially obtain
\beq
\Ait{j}{1}(\qz;\xt) \ns\rightarrow\ns
 : \Ait{j}{1}(\qz;\xt) :.
\eeq
Likewise, since the only 
possible self-contraction which we may consider extracting
from $\Ait{j}{2}(\qz;\xt)$ vanishes,
we have simply
\beq
\Ait{j}{2}(\qz;\xt) \ns\rightarrow\ns
: \Ait{j}{2}(\qz;\xt) :.
\eeq

At third order, in addition to the contribution where we choose
to do no contractions, we have a term which is generated from 
contracting $\rho({\xiz}_2;\pt_2)$ with $\rho({\xiz}_3;\pt_3)$.
The color algebra associated with this contraction is 
straightforward:
\beq
\delta^{ab} \Bl\Bl\rho({\xiz}_1;\pt_1), T^{a} \Br, T^{b} \Br
= N_c \rho({\xiz}_1;\pt_1).
\label{color23}
\eeq
The interesting longitudinal factors read
\beqa
\int_{-\infty}^{\infty} \nts d{\yz}_2
\int_{-\infty}^{\infty} \nts d{\yz}_3
\int_{-\infty}^{\infty} \nts d{\xiz}_2
\int_{-\infty}^{\infty} \nts d{\xiz}_3 \ts &&
\Theta({\yz}_{1}-{\yz}_2) \ts
\Theta({\yz}_{2}-{\yz}_{3}) \ts
\cr \times && 
\exp\Bigl(-p_2 \vert{\yz}_2-{\xiz}_2 \vert \Bigr)
 \exp\Bigl(-p_3 \vert{\yz}_3-{\xiz}_3 \vert \Bigr)
\cr \phantom{\biggl[} \times && 
 {\SIII}\Bigl( \half({\xiz}_2+{\xiz}_3); \pt_2+\pt_3 \Bigr)
 {\DIII}\Bigl({\xiz}_2-{\xiz}_3;\half(\pt_2-\pt_3)\Bigr).
\label{x1528C}
\eeqa
We again make the variable changes indicated in Eqs.~(\ref{vchange1})
and~(\ref{vchange2}), producing
\beqa
\int_{-\infty}^{\infty} \nts d\sigmaz
\int_{-\infty}^{\infty} \nts d\deltaz
\int_{-\infty}^{\infty} \nts d{\uz}_2 &&
\int_{-\infty}^{\infty} \nts d{\uz}_3 \ts 
\Theta({\yz}_1-\sigmaz-\half\deltaz-{\uz}_2) \ts
\Theta(\deltaz+{\uz}_2-{\uz}_3) \ts
\cr \phantom{\biggl[} \times && 
\exp \Bigl( -p_2 \vert {\uz}_2 \vert - p_3 \vert {\uz}_3 \vert \Bigr)
 {\SIII}(\sigmaz;\pt_2+\pt_3)
 {\DIII}\Bigl(\deltaz;\half(\pt_2-\pt_3)\Bigr).
\label{x1528D}
\eeqa
We may apply the large nucleus approximation to drop $\deltaz$
and ${\uz}_2$ relative to $\sigmaz$ in the first $\Theta$-function
appearing in~(\ref{x1528D}).  However, the same arguments which allow
us to do so also tell us that $\deltaz$, ${\uz}_2$, and ${\uz}_3$
are all of order $a$. Hence, the second $\Theta$-function cannot
be simplified.
Nevertheless, dropping ${\uz}_2$ 
and $\deltaz$ from the first $\Theta$-function
is sufficient to allow the
integrations on ${\uz}_2$ and ${\uz}_3$ to proceed easily,
yielding
\beq
{{4}\over{p_2 p_3}}
\int_{-\infty}^{\infty}   d\sigmaz
\ts \Theta({\yz}_1-\sigmaz) 
\ts {\SIII}(\sigmaz;\pt_2+\pt_3)
\int_{0}^{\infty}  d\deltaz \ts
 {\DIII}\Bigl(\deltaz;\half(\pt_2-\pt_3)\Bigr) .
\label{x1613B-1}
\eeq
For a spherically symmetric nucleus, $\DIII$ is an even function
of $\deltaz$.  Thus, we conclude that the $\deltaz$ integration
simply completes the Fourier transform of $\DIII/2$, with the 
longitudinal momentum evaluated at zero:
\beq
{{2}\over{p_2 p_3}}
\ts \widetilde{\DIII}\Bigl(0,\half(\pt_2-\pt_3)\Bigr) 
\int_{-\infty}^{{\yz}_1}  d\sigmaz
\ts {\SIII}(\sigmaz;\pt_2+\pt_3).
\label{x1613B-2}
\eeq
Applying the results in~(\ref{color23}) and~(\ref{x1613B-2}), we find
that
\beqa
\Ait{j}{3}(\qz;\xt) \ns\rightarrow\ns 
: \Ait{j}{3}(\qz;\xt) :  +
{{1}\over{ig\qz}} 
\ts(ig^2) &&
\int {{d^2\pt_1}\over{4\pi^2}} \ts  
{{p_{1j}}\over{2p_1}} \ts
e^{-i\pt_1\cdot\xt} 
\int_{-\infty}^{\infty} d{\yz}_1
\ts \exp(-i\qz{\yz}_1)
\cr \phantom{\Biggl[} \times &&
\int_{-\infty}^{\infty} \nts\nts d{\xiz}_1 \ts
\exp\Bigl(-p_1\vert{\yz}_1 {-} {\xiz}_1\vert\Bigr)
 : \Lbrk
\rho({\xit}_1;\pt_1)
\Rbrk :
\cr \phantom{\Biggl[} \times && 
\Bigl\{ -\half N_c g^4 \Kcube\ts \Gamma({\yz}_1;\xt,\xt)
\Bigr\}.
\label{x1613}
\eeqa
In~(\ref{x1613}) we have introduced the function
\beq
\Gamma(\xz;\xt,\xt') \equiv
\int_{-\infty}^{\xz} d\sigmaz
\int { {d^2\kt}\over{4\pi^2} } 
\int { {d^2\kt'}\over{4\pi^2} } \ts
{
{ e^{-i\kt\cdot\xt} }
\over
{ \kt^2 }
} \ts {
{ e^{-i\kt'\cdot\xt'} }
\over
{ \kt^{\prime 2} }
} \ts 
\SIII(\sigmaz;\kt+\kt') \ts
\widetilde{\DIII}\Bigl(0,\half(\kt-\kt')\Bigr) 
\label{Gamma3}
\eeq
which will prove to be useful as we proceed with the calculation.

At fourth order, there are two different non-vanishing contractions.
Their contributions differ only in the range of the $\sigmaz$
integration, and combine neatly to produce
\beqa
\Ait{j}{4}(\qz;\xt) \ns\rightarrow\ns &&
: \Ait{j}{4}(\qz;\xt) :  \phantom{\Biggl[}
\cr + &&
{{1}\over{ig\qz}} 
(ig^2)^2 
\int {{d^2\pt_1}\over{4\pi^2}}  
\int {{d^2\pt_2}\over{4\pi^2}} \ts
{ { p_{1j} e^{-i\pt_1\cdot\xt} } \over{2p_1}} \ts
{ { e^{-i\pt_2\cdot\xt} } \over{2p_2}} \ts
\int_{-\infty}^{\infty} \nts d^2{\yz}_{\downarrow}
\ts \exp(-i\qz{\yz}_1)
\cr \phantom{\Biggl[} && \qquad\qquad\times 
\int_{-\infty}^{\infty} \nts d^2{\xiz} \ts
\exp\Bigl(-p_1\vert{\yz}_1 {-} {\xiz}_1\vert
          -p_2\vert{\yz}_2 {-} {\xiz}_2\vert \Bigr)
\cr \phantom{\Biggl[} && \qquad\qquad\times
 : \Lbrk
\rho({\xiz}_1;\pt_1)
\rho({\xiz}_2;\pt_2)
\Rbrk :
\Bigl\{ -\half N_c g^4 \Kcube\ts \Gamma({\yz}_1;\xt,\xt)
\Bigr\}.
\eeqa
Finally, at fifth order, in addition to the three different ways
to perform a single contraction, we encounter a contribution 
containing two contractions.  The result of a straightforward 
computation is
\beqa
\Ait{j}{5}(\qz;\xt) \ns\rightarrow\ns &&
: \Ait{j}{5}(\qz;\xt) :  \phantom{\Biggl[}
\cr + &&
{{1}\over{ig\qz}} 
(ig^2)^3 
\int {{d^2\pt_1}\over{4\pi^2}}  
\int {{d^2\pt_2}\over{4\pi^2}} 
\int {{d^2\pt_3}\over{4\pi^3}} \ts
{ { p_{1j} e^{-i\pt_1\cdot\xt} } \over{2p_1}} \ts
{ { e^{-i\pt_2\cdot\xt} } \over{2p_2}} \ts
{ { e^{-i\pt_3\cdot\xt} } \over{2p_3}} \ts
\cr \phantom{\Biggl[} && \quad\qquad\times 
\int_{-\infty}^{\infty} d^3{\yz}_{\downarrow}
\ts \exp(-i\qz{\yz}_1)
\int_{-\infty}^{\infty} d^3{\xiz} \ts
\exp\biggl( -\sum_{\ell=1}^3 p_\ell\vert{\yz}_\ell-{\xiz}_\ell\vert
\biggr)
\cr \phantom{\Biggl[} && \quad\qquad\times
 : \Lbrk
\rho({\xiz}_1;\pt_1)
\rho({\xiz}_2;\pt_2)
\rho({\xiz}_3;\pt_3)
\Rbrk :
\Bigl\{ -\half N_c g^4 \Kcube\ts \Gamma({\yz}_1;\xt,\xt)
\Bigr\}
\cr + &&
{{1}\over{ig\qz}} 
\ts(ig^2)
\int {{d^2\pt_1}\over{4\pi^2}} \ts  
{{p_{1j}}\over{2p_1}} \ts
e^{-i\pt_1\cdot\xt} 
\int_{-\infty}^{\infty} d{\yz}_1
\ts \exp(-i\qz{\yz}_1)
\cr \phantom{\Biggl[} && \quad\qquad\times
\int_{-\infty}^{\infty} d{\xiz}_1 \ts
\exp\Bigl(-p_1\vert{\yz}_1-{\xiz}_1\vert\Bigr)
 : \Lbrk
\rho({\xiz}_1;\pt_1)
\Rbrk :
\cr \phantom{\Biggl[} && \quad\qquad\times
{{1}\over{2!}} \ts
\Bigl\{ -\half N_c g^4 \Kcube\ts \Gamma({\yz}_1;\xt,\xt)
\Bigr\}^2.
\eeqa

At this stage we can see the pattern which is emerging.  When we
choose to do no contractions, we get back the series for $A^j(\qz;\xt)$,
but normal-ordered.  Starting at third order, we have the option of
doing at least one contraction.  Choosing to do exactly one contraction
at each order produces a series which is nearly the one for
$A^j(\qz;\xt)$, but with an extra factor
\beq
-\half N_c g^4 \Kcube\ts \Gamma({\yz}_1;\xt,\xt)
\eeq
inserted into the integrand of each term.
At fifth order, we may elect to do at least two contractions.
Doing exactly two contractions in each term again nearly reproduces
the series for
$A^j(\qz;\xt)$, but this time with an extra factor
\beq
{{1}\over{2!}} \ts
\Bigl\{ -\half N_c g^4 \Kcube\ts \Gamma({\yz}_1;\xt,\xt)
\Bigr\}^2
\eeq
in the integrand.
In like manner, all of the terms in which we do exactly
$j$ contractions sum up to (nearly) produce the series for 
$A^j(\qz;\xt)$, but with the extra factor
\beq
{{1}\over{j!}} \ts
\Bigl\{ -\half N_c g^4 \Kcube\ts \Gamma({\yz}_1;\xt,\xt)
\Bigr\}^j.
\eeq
Thus, we conclude that systematically accounting for all possible 
self-contractions results in
\beqa
A^{j}(\qz;\xt) \enspace\rightarrow\enspace
{{1}\over{ig\qz}} 
\sum_{m=1}^{\infty} 
(ig^2)^{m}  
&& \int {{d^2\pt_1}\over{4\pi^2}} \ts
{ {p_{1j}} \over {2p_1} } \ts
e^{-i\pt_1\cdot\xt} 
\Biggl[ \ts
\prod_{\ell=2}^m
\int {{d^2\pt_\ell}\over{4\pi^2}} \ts
{ {e^{-i\pt_\ell\cdot\xt}} \over {2p_\ell} } 
\Biggr]
\cr \times &&
\int_{-\infty}^{\infty} \nts\nts d^m{\yz}_{\downarrow} 
\ts \exp(-i\qz{\yz}_1)
\int_{-\infty}^{\infty} \nts\nts d^m{\xiz}
\exp\biggl( -\sum_{i=1}^{m} p_i \vert {\yz}_i - {\xiz}_i \vert 
\biggr)
\cr \times && \ts
:\Lbrk
\rho({\xiz}_1;\pt_1)
\rho({\xiz}_2;\pt_2) \cdots 
\rho({\xiz}_m;\pt_m)
\Rbrk: \phantom{\Biggl[}
\cr \times && \ts
\exp\Bigl\{ -\half N_c g^4 \Kcube\ts \Gamma({\yz}_1;\xt,\xt)
\Bigr\}.
\label{x1616}
\eeqa


\subsection{Mutual Contractions}

We now insert the required two copies
of Eq.~(\ref{x1616}) into Eq.~(\ref{gnum3d}),
the formula for the gluon number density. 
Because
all of the self contractions have already been accounted
for, we may 
only multiply terms which contain the same number of $\rho$'s,
leading to a single sum (rather than a double sum).
Thus, we obtain
\beqa
\gnumIII = &&
{ {1}\over{\pi^3g^2} } \ts 
{ {1}\over{\qz} }  
\int d^2\xt \int d^2\xprimet \ts
e^{i\qt\cdot(\xt-\xprimet)}
\sum_{m=1}^{\infty} 
(-g^4)^m  
\nts
\int {{d^2\pt_1}\over{4\pi^2}} 
\int {{d^2\pt_1^\prime}\over{4\pi^2}} \ts
{ {\pt_1\cdot\pt_1^\prime} \over {4 p_1 \ts p_1^\prime} } \ts
e^{-i\pt_1\cdot\xt-i\pt_1^\prime\cdot\xprimet} 
\nts\nts\nts\nts\nts\nts
\cr \times &&
\Biggl[ \ts
\prod_{\ell=2}^m
\int {{d^2\pt_\ell}\over{4\pi^2}} 
\int {{d^2\pt_\ell^\prime}\over{4\pi^2}} \ts
{ {e^{-i\pt_\ell\cdot\xt-i\pt_\ell^\prime\cdot\xprimet}} 
  \over 
  {4 p_\ell\ts p_\ell^\prime} } 
\Biggr]
\int_{-\infty}^{\infty} \nts\nts d^m{\yz}_{\downarrow} 
\int_{-\infty}^{\infty} \nts\nts d^m{\yz}_{\downarrow}^\prime 
\ts \exp\Bigl[-i\qz({\yz}_1 {-} {\yz}_1^\prime)\Bigr]
\cr \times &&
\int_{-\infty}^{\infty} \nts d^m{\xiz}
\int_{-\infty}^{\infty} \nts d^m{\xiz}'
\exp\Biggl\{
- \sum_{i=1}^{m}
\Bigl(
 { p_i \vert {\yz}_i {-} {\xiz}_i \vert } +
 { p_i^\prime \vert {\yz}_i^\prime {-} {\xiz}_i^\prime \vert }
\Bigr)
\Biggr\}
\cr \times &&
\Biggl\langle
\trace \biggl( T^a
{:}\Lbrk
\rho({\xiz}_1;\pt_1)
\cdots 
\rho({\xiz}_m;\pt_m)
\Rbrk{:} 
\biggr)
\trace \biggr( T^a
{:}\Lbrk
\rho({\xiz}_1^\prime;\pt_1^\prime)
\cdots 
\rho({\xiz}_m^\prime;\pt_m^\prime)
\Rbrk{:} 
\biggr)
\Biggr\rangle
\nts\nts\nts\nts\nts\nts
\cr \times &&
\exp\biggl\{ -\half N_c g^4 \Kcube\ts 
\Bigl[\Gamma({\yz}_1;\xt,\xt) 
    + \Gamma({\yz}_1^{\prime};\xt',\xt')\Bigr]
\biggr\}.
\label{GiantMess}
\eeqa
Using an argument exactly analogous to the one in 
Eqs.~(\ref{x1526})--(\ref{x1527a}) it can be shown that the
only non-vanishing contribution to leading order in powers of
$a/R$ is obtained by performing ``corresponding'' contractions
(like the contractions in Fig.~\ref{EighthOrder}d),
that is, $\rho({\xiz}_j;\pt_j)$ with 
$\rho({\xiz}_j^\prime;\pt_j^\prime)$ for all $j$.
``Crossed'' contractions (like Fig.~\ref{EighthOrder}c) are
suppressed by one or more factors of $a/R$.

The color algebra associated with the 
corresponding contractions involves the expression
\beq
{\cal T} \equiv
\trace\biggl(
T^a \Lbrk
T^{i_1}T^{i_2} \cdots T^{i_m} 
\Rbrk
\biggr) \ts
\trace\biggl(
T^a \Lbrk
T^{i_1}T^{i_2} \cdots T^{i_m} 
\Rbrk
\biggr).
\label{x914}
\eeq
The required sums are easily evaluated by beginning with the innermost
commutator:
\beqa
\Bl T^{i_1},T^{i_2} \Br_{\alpha\beta} \ts
\Bl T^{i_1},T^{i_2} \Br_{\gamma\delta} \ts
= &&
if^{i_1 i_2 j} (T^j)_{\alpha\beta} \ts
if^{i_1 i_2 k} (T^k)_{\gamma\delta} 
\cr = &&
-N_c (T^j)_{\alpha\beta} \ts (T^j)_{\gamma\delta} .
\label{x915}
\eeqa
The result of inserting~(\ref{x915}) into~(\ref{x914}) has the
same structure as we started with, but with one less commutator.
Repeating this process until just two color matrices remain in each
trace and doing those traces yields
\beq
{\cal T} = \hbox{${1}\over{4}$} (N_c^2-1)(-N_c)^{m-1}.
\label{x915a}
\eeq

Now let us consider the longitudinal integrations.
The relevant factors are
\beqa
&& \int_{-\infty}^{\infty} \nts d^m{\yz}  
\int_{-\infty}^{\infty} \nts d^m{\yz}^\prime  
\int_{-\infty}^{\infty} \nts d^m{\xiz}
\int_{-\infty}^{\infty} \nts d^m{\xiz}^\prime \ts 
\exp\Bigl[-i\qz({\yz}_1-{\yz}_1^\prime)\Bigr]
\phantom{\Bigl[}
\cr && \ts\times 
\exp\biggl\{
-\sum_{\ell=1}^m \Bigl(
p_\ell\vert{\yz}_\ell {-} {\xiz}_\ell \vert + 
p_\ell'\vert{\yz}_\ell^\prime {-} {\xiz}_\ell^\prime \vert 
\Bigr) \biggr\}
\exp\biggl\{ -\half N_c g^4 \Kcube\ts 
\Bigl[\Gamma({\yz}_1;\xt,\xt) 
    + \Gamma({\yz}_1^{\prime};\xt',\xt')\Bigr]
\biggr\}
\kern-25pt
\cr && \ts\times
\prod_{j=1}^{m}
\Theta({\yz}_{j-1} {-} {\yz}_j) \ts
\Theta({\yz}_{j-1}^\prime {-} {\yz}_j^\prime) \ts
 {\SIII}\Bigl(\half({\xiz}_j {+} {\xiz}_j^\prime);
              \pt_j {+} \pt_j'\Bigr) \ts
 {\DIII}\Bigl({\xiz}_j {-} {\xiz}_j^\prime;
              \half(\pt_j {-} \pt_j^\prime)\Bigr),
\label{x1528N}
\eeqa
where we have defined 
${\yz}_0 = {\yz}_0^\prime = \infty$ for convenience.  
We perform variable changes which are completely analogous 
to those in Eqs.~(\ref{vchange1}) and~(\ref{vchange2}) and
once again apply the large nucleus ($R\gg a$) approximation:
\beqa
\int_{-\infty}^{\infty} && \nts\nts d^m\sigmaz  
\int_{-\infty}^{\infty} \nts\nts d^m\deltaz
\int_{-\infty}^{\infty} \nts\nts d^m \uz
\int_{-\infty}^{\infty} \nts\nts d^m \uz^\prime \ts 
\exp\Bigl[-i\qz({\deltaz}_1{+}{\uz}_1{-}{\uz}_1^\prime)\Bigr]
\cr \phantom{\Bigl[} && \times
\exp\biggl\{
-\sum_{\ell=1}^{m}
\Bigl( p_\ell\vert {\uz}_\ell \vert 
      + p_\ell^\prime \vert {\uz}_\ell^\prime \vert \Bigr)
\biggr\} \ts
\exp\biggl\{ -\half N_c g^4 \Kcube\ts 
\Bigl[\Gamma({\sigmaz}_1;\xt,\xt) 
    + \Gamma({\sigmaz}_1;\xt',\xt')\Bigr]
\biggr\} 
\cr \phantom{\Bigl[} && \times
\prod_{j=1}^{m}
\Theta({\sigmaz}_{j-1} {-} {\sigmaz}_j) \ts
 {\SIII}( {\sigmaz}_j ; \pt_j {+} \pt_j^\prime)  \ts
{\DIII}({\deltaz}_j;\half(\pt_j {-} \pt_j^\prime)\Bigr).
\label{x1528O}
\eeqa
The necessary $\uz$ and $\uz'$ integrations are all easily
performed using
\beq
\int_{-\infty}^{\infty} \nts\nts d\uz
\ts\exp\Bigl( -i\qz\uz - p\vert\uz\vert \Bigr ) = 
{ {2p}\over{\pt^2 + \qz^2} }.
\eeq
The $\delta$ integrals simply finish Fourier-transforming $\DIII$
on the longitudinal coordinate:  
\beq
\int_{-\infty}^{\infty} \nts d{\deltaz} \ts
\exp(-i\qz\deltaz)\ts
\DIII\Bigl({\deltaz}; \half(\pt-\pt^\prime)\Bigr)
= \widetilde\DIII\Bigl(\qz,\half(\pt-\pt^\prime)\Bigr).
\eeq
Applying these considerations to~(\ref{x1528O}) produces
\beqa
\int_{-\infty}^{\infty} \nts\nts d{\sigmaz}_1  &&
{ {2p}_1\over{\pt^2_1 + \qz^2} } \ts
{ {2p'_1}\over{\pt^{\prime 2}_1 + \qz^2} } \ts
{\SIII}( {\sigmaz}_1 ; \pt_1 {+} \pt_1^\prime)  \ts
\widetilde{\DIII}\Bigl(\qz,\half(\pt_1-\pt'_1)\Bigr) 
\cr \phantom{\Bigl[} && \qquad\times
\exp\biggl\{ -\half N_c g^4 \Kcube\ts 
\Bigl[\Gamma({\sigmaz}_1;\xt,\xt) 
    + \Gamma({\sigmaz}_1;\xt',\xt')\Bigr]
\biggr\} 
\cr \phantom{\Bigl[} && \qquad\times
\int_{-\infty}^{{\sigmaz}_1} \nts\nts d^{m-1}{\sigmaz}_{\downarrow}
\prod_{j=2}^{m}
{{4}\over{p_j \ts p'_j}} \ts
{\SIII}( {\sigmaz}_j ; \pt_j {+} \pt_j^\prime)  \ts
\widetilde{\DIII}\Bigl(0,\half(\pt_j {-} \pt_j^\prime)\Bigr).
\label{AnExtraStep}
\eeqa
Notice that
the $\sigmaz$ integrations have inherited the ordering associated
with the gauge transformation.  
When we insert~(\ref{AnExtraStep}) back into Eq.~(\ref{GiantMess}),
the portion of the resulting ordered integrand involving
${\sigmaz}_2,\ldots,{\sigmaz}_m$ is symmetric, allowing us to 
do the sum on $m$ to obtain an exponential.
Including the color factor contained in Eq.~(\ref{x915a})
we arrive at
\beqa
\gnumIII = &&
\Kcube (N_c^2{-}1) \ts
{ { \alpha_s } \over{\pi^2} } \ts 
{ {1}\over{\qz} }  
\int d^2\xt \int d^2\xprimet \ts
e^{i\qt\cdot(\xt-\xprimet)}
\cr \times &&
\int_{-\infty}^{\infty} \nts\nts d{\sigmaz}_{1} 
\int {{d^2\pt_1}\over{4\pi^2}} 
\int {{d^2\pt_1^\prime}\over{4\pi^2}} \ts
{ 
{(-\pt_1\cdot\pt'_1) e^{-i\pt_1\cdot\xt-i\pt_1^\prime\cdot\xprimet} } 
\over{(\pt^2_1 + \qz^2) (\pt^{\prime 2}_1 + \qz^2)} 
} \ts
{\SIII}( {\sigmaz}_1 ; \pt_1 {+} \pt_1^\prime)  \ts
\widetilde{\DIII}\Bigl(\qz,\half(\pt_1 {-} \pt'_1)\Bigr) 
\kern-30pt
\cr && \qquad\times 
\exp\biggl\{ N_c g^4 \Kcube\ts 
\Bigl[ \Gamma({\sigmaz}_1;\xt,\xt')
      -\half\Gamma({\sigmaz}_1;\xt,\xt) 
      -\half\Gamma({\sigmaz}_1;\xt',\xt')\Bigr]
\biggr\}.
\label{x1617}
\eeqa

To proceed further requires us to apply the consequences of the
large nucleus approximation to the transverse coordinates.
To see how this works, let us examine the function $\Gamma$
a bit more closely.  From Eq.~(\ref{Gamma3}) we may write
\beq
\Gamma({\sigmaz}_1;\xt,\xt') =
\int_{-\infty}^{{\sigmaz}_1} d\sigmaz
\int { {d^2\qt}\over{4\pi^2} } \ts
 e^{-i\qt\cdot(\xt+\xprimet)/2} 
\SIII(\sigmaz;\qt) 
\int { {d^2\pt}\over{4\pi^2} }  \ts
{
{ e^{-i\pt\cdot(\xt-\xprimet)} \widetilde{\DIII}(0,\pt) }
\over
{ (\pt+\half\qt)^2 \ts (\pt-\half\qt)^2 }
} 
\label{GammaPQ}
\eeq
where we have changed variables to $\qt\equiv \kt{+}\kt'$
and $\pt \equiv \half(\kt{-}\kt')$.
Recall from the discussion in the paragraph following Eq.~(\ref{x1527})
that the values of the momenta associated with $\SIII$
are $q \sim 1/R$ whereas those associated with $\DIII$ are
$p \sim 1/a$.  
This suggests that we may neglect $\qt$ in the two denominators
of Eq.~(\ref{GammaPQ}), the error being suppressed by
a factor of $a/R$.  However, we must be careful.  The combination
appearing in the square brackets of Eq.~(\ref{x1617}) can be shown
to be infrared finite provided that $\widetilde{\DIII}$ is
rotationally invariant and satisfies the color neutrality condition.
This is true to all orders in $a/R$.   When dropping 
terms which are higher order in $a/R$, we should avoid introducing an
infrared divergence, since none was present in the original expression.
Therefore we write
\beqa
\Gamma({\sigmaz}_1;&&\xt,\xt')
   -\half\Gamma({\sigmaz}_1;\xt,\xt) 
   -\half\Gamma({\sigmaz}_1;\xt',\xt') =
\phantom{\biggl[}
\cr &&
\int_{-\infty}^{{\sigmaz}_1} d\sigmaz
\int { {d^2\qt}\over{4\pi^2} } \ts
 e^{-i\qt\cdot(\xt+\xprimet)/2} 
\SIII(\sigmaz;\qt) 
\int { {d^2\pt}\over{4\pi^2} }  \ts
{
{ \widetilde{\DIII}(0,\pt) }
\over
{ \pt^4 }
} 
\ts \Bigl[
e^{-i\pt\cdot(\xt-\xprimet)}-1
\Bigr] 
+ {\cal O}\biggl({{a}\over{R}}\biggr),
\label{Replacement}
\eeqa
that is, when we drop $\qt$ from the denominators we should 
simultaneously adjust the exponential multiplying $\SIII$ to be
identical in all three terms.
The advantage of the form contained in~(\ref{Replacement}) is
the decoupling of the two momentum integrations.  The integral
on $\qt$ just converts $\SIII(\sigmaz;\qt)$ back to a purely
position-space quantity.  The $\pt$ integral defines the function
\beq
L(\xt)  \equiv
\int { {d^2\pt}\over{4\pi^2} } \ts
{
{ \widetilde{\DIII}(0,\pt) }
\over
{  \pt^4  }
}
\ts \Bigl[
e^{-i\pt\cdot\xt}-1
\Bigr] .
\label{x1488a-APPX}
\eeq
Thus,
\beqa
\Gamma({\sigmaz}_1;\xt,\xt')
   -\half\Gamma({\sigmaz}_1;\xt,\xt) 
   -\half&&\Gamma({\sigmaz}_1;\xt',\xt') =
\phantom{\biggl[}
\cr &&
L(\xt-\xt')
\int_{-\infty}^{{\sigmaz}_1} d\sigmaz \ts\ts
\SIII\Bigl(\sigmaz,\half(\xt+\xprimet)\Bigr) 
+ {\cal O}\biggl({{a}\over{R}}\biggr).
\label{Replacement2}
\eeqa
Treating the $\pt_1$ and $\pt'_1$ integrals of Eq.~(\ref{x1617})
in a similar manner and applying Eq.~(\ref{Replacement2})
yields the expression
\beqa
\gnumIII = 
\Kcube (N_c^2{-}1) \ts
{ { 2\alpha_s } \over{\pi^2} } \ts 
{ {1}\over{\qz} }  
\int && d^2\xt  \int d^2\xprimet \ts
e^{i\qt\cdot(\xt-\xprimet)}
\int_{-\infty}^{\infty} \nts\nts d{\sigmaz}_{1} 
\curlyL(\qz;\xt-\xt') \ts
{\SIII}\Bigl( {\sigmaz}_1 , \half(\xt{+}\xt')\Bigr)  
\kern-25pt
\cr && \times 
\exp\biggl\{ N_c g^4 \Kcube L(\xt{-}\xt')
\int_{-\infty}^{{\sigmaz}_1} d\sigmaz \ts\ts
\SIII\Bigl(\sigmaz,\half(\xt+\xprimet)\Bigr) 
\biggr\},
\label{x1617-b}
\eeqa
where we have introduced the quantity
\beq
\curlyL(\qz;\xt) \equiv {{1}\over{2}}
\int { {d^2\pt}\over{4\pi^2} } \ts
e^{-i\pt\cdot\xt} \ts
{
{ \pt^2 \ts \widetilde{\DIII}(\qz,\pt) }
\over
{ ( \pt^2 + \qz^2 )^2 }
}.
\label{x1490-APPX}
\eeq
Finally, we apply the chain rule to do the integral over
${\sigmaz}_1$, and switch to sum and difference variables for the
$\xt$ and $\xprimet$ integrations:
\beqa
\gnumIII = 
\Kcube(N_c^2-1) \ts
{ { 2\alpha_s } \over{\pi^2} } \ts 
{ {1}\over{\qz} }  &&
\int d^2\deltat \ts
e^{i\qt\cdot\deltat} \curlyL(\qz;\deltat) \ts
\cr \times &&
\int d^2\sigmat \ts
{
{ \exp\Bigl\{ g^4 N_c \Kcube L(\deltat)
  \int_{-\infty}^{\infty} d\sigmaz \ts\ts
  \SIII(\sigmaz,\sigmat)\Bigr\} - 1 }
\over
{ g^4 N_c \Kcube L(\deltat) }
}.
\label{x1622-APPX}
\eeqa


\subsection{Geometric Dependence}

In order to perform the $\sigmat$ integration appearing in
Eq.~(\ref{x1622-APPX}), it is necessary to specify the geometry
of the nucleus.  We will consider two cases, cylindrical and
spherical.

A cylindrical nucleus is described by the function
\beq
\SIII(\sigmaIII\ts) = \Theta(R^2 - \sigmat^2)  \ts
                  \Theta\Bigl((\half h)^2 - \sigmaz^2 \Bigr),
\eeq
where $R$ is the radius of the cylinder and $h$ is its height.
Actually, the height will drop out of the final result, 
since~(\ref{x1622-APPX}) depends on
\beqa
\Kcube\int_{-\infty}^{\infty} d\sigmaz \ts
    \SIII(\sigmaz,\sigmat) 
    &=& \Kcube\ts h \ts \Theta(R^2 - \sigmat^2)
\cr &=&
    {{3A}\over{2N_c}} \ts {{1}\over{\pi R^2}} \ts \Theta(R^2-\sigmat^2).
\label{NOh}
\eeqa
The $\sigmat$ integral which results from inserting~(\ref{NOh})
into~(\ref{x1622-APPX}) is trivial, producing
\beqa
\gnumIII = 
3AC_F \ts
{ { 2\alpha_s } \over{\pi^2} } \ts 
{ {1}\over{\qz} }  &&
\int{  d^2\deltat \ts
e^{i\qt\cdot\deltat} \curlyL(\qz;\deltat) \ts
{
{ \exp\Bigl\{ [{3Ag^4}/{2\pi R^2}] \ts L(\deltat)\Bigr\} - 1 }
\over
{ [{3Ag^4}/{2\pi R^2}] \ts L(\deltat) }
}},
\label{x1628}
\eeqa
which is equivalent to the portions of 
Eqs.~(\ref{x1629})--(\ref{x1629V}) pertaining to cylindrical
geometry.

Turning to the more realistic case of a spherical nucleus, we
have
\beqa
\int_{-\infty}^{\infty} d\sigmaz\ts \SIII(\sigmaz,\sigmat) &=& 
\int_{-\infty}^{\infty} d\sigmaz\ts \Theta(R^2-\sigmat^2-\sigmaz^2)
\cr &=& 2\sqrt{R^2-\sigmat^2} \ts \Theta(R^2-\sigmat^2),
\phantom{\Biggl[}
\eeqa
so that Eq.~(\ref{x1622-APPX}) becomes
\beqa
\gnumIII = 
\Kcube(N_c^2-1) \ts
{ { 2\alpha_s } \over{\pi^2} } \ts 
{ {1}\over{\qz} }  &&
\int d^2\deltat \ts
e^{i\qt\cdot\deltat} \curlyL(\qz;\deltat) \ts
\cr \times &&
\int d^2\sigmat \ts
{
{ \exp\Bigl\{ 2 g^4 N_c \Kcube L(\deltat) 
  \sqrt{R^2-\sigmat^2} \ts \Theta(R^2-\sigmat^2)\Bigr\} - 1 }
\over
{ g^4 N_c \Kcube L(\deltat) }
}.
\label{x1623top}
\eeqa
Thus, the $\sigmat$ integral hinges upon the form
\beq
\zed \equiv \int_0^R d\Sigma\ts\Sigma\ts 
\Biggl[\exp\Bigl( \Omega\sqrt{R^2-\Sigma^2} \ts \Bigr) - 1\biggr].
\eeq
This integral is easily performed by the change of variables
\beq
s = \Omega \sqrt{R^2-\Sigma^2}; 
\qquad s\ts ds = -\Omega^2 \Sigma \ts d\Sigma.
\eeq
Then
\beqa
\zed &=& {{1}\over{\Omega^2}} 
\int_{0}^{\Omega R} ds \ts s(e^s-1) \phantom{\Biggl[}
\cr &=&
{{1}\over{\Omega^2}}\ts
\Bigl[ 1 - \half(\Omega R)^2 + e^{\Omega R} (\Omega R - 1) \Bigr].
\label{x1626A}
\eeqa
Applying~(\ref{x1626A}) to~(\ref{x1623top}) leads to the result
\beqa
\gnumIII = 
3AC_F \ts
{ { 2\alpha_s } \over{\pi^2} } \ts 
{ {1}\over{\qz} }  &&
\int d^2\deltat \ts
e^{i\qt\cdot\deltat} \curlyL(\qz;\deltat) \ts
\cr && \times
{{3}\over{[\vsqr L(\deltat)]^3}} \ts
\Biggl\{ 1 - \half[\vsqr L(\deltat)]^2
+ \Bigl[ \vsqr L(\deltat) - 1 \Bigr] \exp\Bigl[\vsqr L(\deltat) \Bigr]
\Biggr\},
\label{x1629-APPX}
\eeqa
where $\vsqr = 9A g^4 / 4\pi R^2 $.


\section{Calculational Details for the Power-Law Model}\label{PowerDetails}


\subsection{Useful Integrals}

All of the integrals required to compute the 
functions $\curlyL(\qz;\xt)$ and $L(\xt)$
which appear in the integrand for the gluon number density
for the power-law (Yukawa-like) model introduced in
Sec.~\ref{EXAMPLE} may be derived from the forms
\beq
{\cal I}_{1\omega} \equiv
\int
{ {d^2\qt}\over{4\pi^2} } \ts
{
{ e^{-i\qt\cdot\xt} }
\over
{ (\qt^2+\qz^2) \ts [1+b^2(\qt^2+\qz^2)]^\omega }
},
\label{Ij1}
\eeq
and
\beq
{\cal I}_{2\omega} \equiv
\int
{ {d^2\qt}\over{4\pi^2} } \ts
{
{ e^{-i\qt\cdot\xt} }
\over
{ (\qt^2+\qz^2)^2 \ts [1+b^2(\qt^2+\qz^2)]^\omega }
}.
\label{Ij2}
\eeq
Since the $\omega \rightarrow 0$ limits of~(\ref{Ij1}) and~(\ref{Ij2})
are smooth, we may simply set $\omega=0$  to obtain the
necessary single-denominator integrals.
Because the procedure for performing both
integrals is essentially the same, we will describe the computation
for ${\cal I}_{2\omega}$ and simply quote the result for 
${\cal I}_{1\omega}$.

The computation of ${\cal I}_{2\omega}$ begins by 
introducing a single Feynman parameter to combine the two
denominators:
\beq
{\cal I}_{2\omega} = 
{ {\omega (\omega {+}1)}\over{b^{2\omega }} }
\int_0^1 dz \ts z^{\omega -1}(1{-}z)
\int
{ {d^2\qt}\over{4\pi^2} } \ts
{
{ e^{-i\qt\cdot\xt} }
\over
{ (\qt^2 + \qz^2 + z/b^2)^{\omega+2} }
}.
\eeq
In order to deal with the $\qt$ integration, we
introduce a Schwinger parameter to promote the denominator
into the exponential:
\beq
{\cal I}_{2\omega} =
{ {1}\over{b^{2\omega}} } \ts
{ {1}\over{(\omega-1)!} }
\int_0^{\infty} d\lambda \ts \lambda^{\omega+1}
\int_0^1 dz \ts z^{\omega -1}(1{-}z)
\int
{ {d^2\qt}\over{4\pi^2} } \ts
e^{-i\qt\cdot\xt} e^{-\lambda(\qt^2+\qz^2 + z/b^2)}.
\eeq
The $\qt$ integration is now Gaussian, and may be performed in the
usual fashion, with the result
\beq
{\cal I}_{2\omega} = 
{ {1}\over{4\pi b^{2\omega}} } \ts
{ {1}\over{(\omega-1)!} }
\int_0^{\infty} d\lambda \ts \lambda^{\omega}
\exp\biggl[
-{{1}\over{\lambda}} \ts { {\xt^2}\over{4} }
- \lambda \qz^2
\biggr]
\int_0^1 dz \ts z^{\omega-1} (1{-}z) \ts e^{-z\lambda/b^2}.
\eeq
The $z$ integration is straightforward, yielding
\beqa
{\cal I}_{2\omega} &=& 
{ {1}\over{4\pi} } 
\int_0^{\infty} 
d\lambda \ts
\Biggl[ 1 - e^{-\lambda/b^2} \sum_{\ell=0}^{\omega-1}
{ {1}\over{\ell!} }
\biggl( {{\lambda}\over{b^2}} \biggr)^\ell \ts
\Biggr] \ts
\exp\biggl[
-{{1}\over{\lambda}} \ts { {\xt^2}\over{4} }
- \lambda \qz^2
\biggr]  
\cr &-&
{ {1}\over{4\pi} }  \ts \omega b^2
\int_0^{\infty} 
{{d\lambda}\over{\lambda}} \ts
\Biggl[ 1 - e^{-\lambda/b^2} \sum_{\ell=0}^{\omega}
{ {1}\over{\ell!} }
\biggl( {{\lambda}\over{b^2}} \biggr)^\ell \ts
\Biggr] \ts
\exp\biggl[
-{{1}\over{\lambda}} \ts { {\xt^2}\over{4} }
- \lambda \qz^2
\biggr].  
\eeqa
It is convenient at this stage to introduce the dimensionless
integration variable $\xi \equiv \lambda/b^2$.  Doing so and
performing a bit of algebra we arrive at
\beqa
{\cal I}_{2\omega} &=& 
{{b^2}\over{4\pi^2}} 
\int_0^{\infty} d\xi \ts
\exp\biggl[
-{{1}\over{\xi}} \ts \biggl( {{x}\over{2b}} \biggr)^2
- \xi (b\qz)^2
\biggr]
\phantom{\Biggl[} \cr &-&
{{b^2}\over{4\pi^2}} \ts \omega
\int_0^{\infty} {{d\xi}\over{\xi}} \ts
\exp\biggl[
-{{1}\over{\xi}} \ts \biggl( {{x}\over{2b}} \biggr)^2
- \xi (b\qz)^2
\biggr]
\phantom{\Biggl[} \cr &+&
{{b^2}\over{4\pi^2}} \ts 
\sum_{\ell=0}^{\omega-1}
{{\omega{-}\ell}\over{\ell !}}\ts
\int_0^{\infty} d\xi \ts
\xi^{\ell-1} \ts
\exp\biggl\{
-{{1}\over{\xi}} \ts \biggl( {{x}\over{2b}} \biggr)^2
- \xi \Bigl[1+(b\qz)^2\Bigr]
\biggr\}.
\eeqa
The $\xi$ integrals may be performed 
to produce modified Bessel functions, as seen
from Eq.~(3.471.9)
of Ref.~\cite{PhysicistsFriend}:
\beq
\int_0^{\infty} dy \ts y^{\mu-1} 
\exp\biggl( -{{\beta}\over{y}} - \gamma y \biggr)
= 2 \biggl( {{\beta}\over{\gamma}} \biggr)^{\mu/2}
K_{\mu}\Bigl(2\sqrt{\beta\gamma}\ts\Bigr),
\label{3.471.9}
\eeq
which is valid for all values of $\mu$, provided that $\beta$ and
$\gamma$ are positive.  Thus, we arrive at
\beqa
{\cal I}_{2\omega} =
{{b^2}\over{2\pi}} \ts 
\Biggl\{ &&
{{x}\over{2b}} \ts
{{1}\over{b\qz}} \ts
K_1(x\qz) - \omega K_0(x\qz)
\cr &&
+ \sum_{\ell=0}^{\omega-1}
{{\omega{-}\ell}\over{\ell !}} \ts
\biggl(
{ {x}\over{2b} } 
\biggr)^{\ell}
\Bigl[ 1 + (b\qz)^2 \Bigr]^{-\ell/2}
K_{\ell} \biggl( {{x}\over{b}}
\sqrt{ 1 + (b\qz)^2 } \ts\biggr)
\Biggr\}.
\label{Ij2solved}
\eeqa
The analogous treatment of~(\ref{Ij1}) yields
\beq
{\cal I}_{1\omega} = 
{ {1}\over{2\pi} }  \ts
\Biggl\{
K_0(x \qz)
- \sum_{\ell=0}^{\omega-1}
{{1}\over{\ell !}} \ts
\biggl(
{ {x}\over{2b} } 
\biggr)^{\ell}
\Bigl[ 1 + (b\qz)^2 \Bigr]^{-\ell/2}
K_{\ell} \biggl( {{x}\over{b}}
\sqrt{ 1 + (b\qz)^2 } \ts\biggr)
\Biggr\}.
\label{Ij1solved}
\eeq

\subsection{Computation of $\curlyL(\qz;\xt)$ and $L(\xt)$}

Eq.~(\ref{x1490}) defines the function which governs
the Abelian portion
of the integrand for the gluon number density.
For the power law model, it is helpful to rewrite the
numerator using
$\pt^2 \equiv \pt^2+\qz^2 - \qz^2$:
\beq
\curlyL(\qz;\xt) = {{1}\over{2}}
\int { {d^2\pt}\over{4\pi^2} } \ts
e^{-i\pt\cdot\xt} \ts
\Biggl[
{ {1}\over{ \pt^2 + \qz^2 } }
-
{ {\qz^2}\over{ (\pt^2 + \qz^2)^2 } }
\Biggr]
\Biggl[
1 - { {1}\over{[1+\aw^2(\pt^2+\qz^2)]^{\omega}} }
\Biggr].
\eeq
In terms of the integrals~(\ref{Ij1})
and~(\ref{Ij2}) introduced in the first part of this
appendix, we have simply
\beq
\curlyL(\qz;\xt) = \half {\cal I}_{10}
                 - \half \qz^2 {\cal I}_{20}
                 - \half {\cal I}_{1\omega}
                 + \half \qz^2 {\cal I}_{2\omega}.
\eeq
A straightforward substitution of the results contained
in Eqs.~(\ref{Ij2solved}) and~(\ref{Ij1solved}) leads to
\beqa
\curlyL(\qz;\xt) =
&-& {{\omega}\over{4\pi}}\ts (\aw\qz)^2 K_0(x \qz)
\phantom{\Biggl[}
\cr &+&
{{1}\over{4\pi}}
\sum_{j=0}^{\omega-1}
{{1}\over{j!}} \ts
\biggl( {{x}\over{2\aw}} \biggr)^j \ts
{
{  1 + (\omega{-}j)(\aw\qz)^2  }
\over
{ \Bigl[ 1 + (\aw\qz)^2 \Bigr]^{j/2} }
} \ts
K_{j} \biggl( {{x}\over{\aw}}
\sqrt{ 1 + (\aw\qz)^2 } \ts\biggr).
\label{x1648}
\eeqa

The determination of $L(\xt)$ from Eq.~(\ref{x1488a}) is a
bit more involved.  The difficulty lies in the fact that it
is not possible to integrate~(\ref{x1488a}) term-by-term,
as the individual bits are infrared divergent.  To work
around this difficulty, let us define the auxiliary function
\beq
L_3(\qz;\xt) \equiv
\int { {d^2\pt}\over{4\pi^2} } \ts
{ {e^{-i\pt\cdot\xt}} \over { (\pt^2 + \qz^2)^2 } }
\ts \widetilde\DIII(\qz,\pt)  .
\label{x1563L}
\eeq
Then, the integral we seek may be determined from
the relation
\beq
L(\xt) = 
\lim_{\qz\rightarrow 0}
\Bigl[ L_3(\qz;\xt) - L_3(\qz;\zero) \Bigr].
\label{x1562}
\eeq
The computation of $L_3(\qz;\xt)$ is straightforward:
inserting the power-law form of $\widetilde\DIII$ given
in Eq.~(\ref{PowerLaw}) we see that
\beq
L_3(\qz;\xt) = {\cal I}_{20} - {\cal I}_{2\omega}.
\eeq
Thus, the application of Eq.~(\ref{Ij2solved}) gives
\beqa
L_3(\qz;\xt) =
&-&{{\aw^2}\over{2\pi}} \ts \omega
\Biggl[
K_0 \biggl( {{x}\over{\aw}}
\sqrt{ 1 + (\aw\qz)^2 } \ts\biggr)
- K_0(x\qz)
\Biggr]
\cr &-&
{{\aw^2}\over{2\pi}} 
\sum_{\ell=1}^{\omega-1}
{{\omega{-}\ell}\over{\ell !}} \ts
\biggl(
{ {x}\over{2\aw} } 
\biggr)^{\ell}
\Bigl[ 1 + (\aw\qz)^2 \Bigr]^{-\ell/2}
K_{\ell} \biggl( {{x}\over{\aw}}
\sqrt{ 1 + (\aw\qz)^2 } \ts\biggr).
\label{x1649}
\eeqa
To determine $L_3(\qz;\zero)$ from~(\ref{x1649}) we require
the following forms of the modified Bessel
functions for small values of the argument:
\beq
K_{\mu}(z) \ts
\mathop{\longrightarrow}_{\scriptstyle{z\rightarrow 0}} \ts
\cases{ 
-  \ln\biggl( \displaystyle{{z}\over{2}} \biggr) - \gamma_E,
                       & $\mu = 0$;\cr
\phantom{x} & \cr
\displaystyle{ {\Gamma(\mu)}\over{2} } \ts
\biggl( \displaystyle{ {2}\over{z} }\biggr)^\mu ,
                       & $\mu\ne 0$, \cr}
\label{Jackson}
\eeq
where $\gamma_E$ is Euler's constant.
Consequently,
\beq
L_3(\qz;\zero) = 
{{\aw^2}\over{4\pi}}
\ts \omega
\ln\Biggl[
{  {1 + (\aw\qz)^2}
   \over
   {(\aw\qz)^2}  }
\Biggr]
- {{\aw^2}\over{4\pi}}
\sum_{\ell=1}^{\omega-1}
{{\omega-\ell}\over{\ell}} \ts
{ {1} \over {[1+(\aw\qz)^2]^\ell} }.
\label{x1650}
\eeq
Subtracting~(\ref{x1650}) from~(\ref{x1649}) and taking the
$\qz\rightarrow 0$ limit gives the final result
\beqa
L(\xt) =
&-& {{\aw^2}\over{2\pi}}\ts\omega
\Biggl[ K_0 \biggl( {{x}\over{\aw}} \biggr)
+ \ln\biggl( {{x}\over{2\aw}} \biggr)
+ \gamma_E \ts
\Biggr]
\cr &+&
{{\aw^2}\over{2\pi}} 
\sum_{\ell =1}^{\omega-1}
(\omega{-}\ell) \ts
\Biggl[
{{1}\over{2\ell}} -
{{1}\over{\ell !}} \ts
\biggl( {{x}\over{2\aw}} \biggr)^\ell \ts
K_{\ell} \biggl( {{x}\over{\aw}} \biggr)
\Biggr].
\label{x1651}
\eeqa




\end{document}